\documentclass{article} \usepackage{graphicx,emulateapj}

\submitted{} \lefthead{BIALEK, EVRARD \& MOHR} \righthead{
Effects of Preheating in X--ray Clusters }

\def\myputfigure#1#2#3#4#5
{\hskip0.03\textwidth\vskip#5pt
\makebox[0pt]{\hskip#2in
\includegraphics[width=#3\textwidth]{#1}}\vskip#4pt\hfill}

\def \dc   {{\Delta_c}}

\def \micm {\hbox{$M_{ICM}$}}

\def \sig8 {\hbox{$\sigma_8$} }
\def \sz   {Sunyaev--Zel'dovich effect }

\def \ficmhat {\hbox{$\hat{f}_{ICM}$}}
\def \ghat {\hbox{$\hat{g}$}}
\def \kevcmsq {\hbox{$\,${\rm keV~cm}$^2$}}
\def \chandra {\hbox{\sl Chandra\/}}

\def \kpc       {{\rm\ kpc}}

\def \yr        {{\rm\ yr}}

\def \eg        {\hbox{\it e.g.} }
\def \seg       {\hbox{\it e.g.}}
\def \cf        {\hbox{\it cf.} }
\def \vs        {\hbox{\it vs.} }
\def \etal      {et al.\ }

\def \K         { \hbox{$\,$ K} }

\def \kmsmpc    {{\rm\ km\ s^{-1}\ Mpc^{-1}}}
\def \kev       {{\rm\ keV}}
\def \msol      {{\rm\ M}_\odot}

\def \hinv      {\hbox{$\, h^{-1}$} }

\def \se        {\!=\!}

\def \ssim      {\! \sim \!}
\def \ssimeq    {\! \simeq \!}
\def \sequiv    {\! \equiv \!}
\def \spropto   {\! \propto \!}
\def\\{\hfil\break}
\def\spose#1{\hbox to 0pt{#1\hss}}
\def\lta{\mathrel{\spose{\lower 3pt\hbox{$\mathchar"218$}}
     \raise 2.0pt\hbox{$\mathchar"13C$}}}
\def\gta{\mathrel{\spose{\lower 3pt\hbox{$\mathchar"218$}}
     \raise 2.0pt\hbox{$\mathchar"13E$}}}

\def\lesssim{\mathrel{\hbox{\rlap{\hbox{\lower4pt\hbox{$\sim$}}}\hbox{$<$}}}}
\def\gtrsim{\mathrel{\hbox{\rlap{\hbox{\lower4pt\hbox{$\sim$}}}\hbox{$>$}}}}

\newcount\itemno
\def \ino         { \the\itemno\global\advance\itemno by 1 }
\def\apj{ApJ}

\def\apjs{ApJS}
\def\aa{A\&A}
\def\mnras{MNRAS}

\def\pasj{PASJ}

\def \xray {\hbox{X--ray} }
\def \rfiveh {\hbox{$r_{500}$}}

\def \lxt {\hbox{$L_X$--$T$} }
\def \slxt {\hbox{$L_X$--$T$}}
\def \mt {\hbox{$M_{ICM}$--$T$} }
\def \smt {\hbox{$M_{ICM}$--$T$}}

\def \st {\hbox{$R_I$--$T$} }
\def \sst {\hbox{$R_I$--$T$}}

\def \logTd6 {\hbox{log$( T/6 \kev)$} }

\begin{document}

\title{Effects of Preheating on X-ray Scaling Relations in Galaxy Clusters}
\author{John J. Bialek\altaffilmark{1}, August
E. Evrard\altaffilmark{1,2} \& Joseph J. Mohr\altaffilmark{3,4,5}}

\altaffiltext{1}{Department of Physics, 1049 Randall Lab, University of Michigan, Ann Arbor, MI 48109} 
\altaffiltext{2}{Department of Astronomy, Dennison Building, University of Michigan, Ann Arbor, MI 48109} 
\altaffiltext{3}{Chandra Fellow} 
\altaffiltext{4}{Departments of Astronomy and Physics,
University of Illinois, 1002 W. Green St, Urbana, IL 61801}
\altaffiltext{5}{Department of Astronomy and Astrophysics, 5640 S 
Ellis Ave, University of Chicago, Chicago, IL, 60637}

\authoremail{jbialek@umich.edu}
\authoremail{evrard@umich.edu}
\authoremail{jmohr@astro.uiuc.edu}

\begin{abstract}
The failure of purely gravitational and gas dynamical models of
\xray cluster formation to reproduce basic observed properties of
the local cluster population suggests the need for one or more
additional physical processes operating on the intracluster
medium (ICM).  We present results from 84 moderate resolution gas
dynamic simulations designed to systematically investigate the
effects of preheating --- an early elevated ICM adiabat ---
on the resultant, local \xray size--temperature,
luminosity--temperature and ICM mass--temperature relations.
Seven sets of twelve simulations are performed for a $\Lambda$CDM
cosmology, each set characterized by a different initial entropy level $S_i$.
The slopes of the observable relations steepen monotonically 
as $S_i$ is increased.  Observed slopes for all three
relations are reproduced by models with $S_i \in 55-150 \kevcmsq$, 
levels that compare favorably to empirical determinations of core ICM
entropy by Lloyd-Davies, Ponman \& Cannon.  
The redshift evolution for the case of a locally successful model with
$S_i \se 106 \kevcmsq$ is presented.  At temperatures $kT\gta
3$~keV, little or no evolution in physical isophotal sizes or bolometric
luminosities is expected to $z \lta 1$.  The ICM mass and total
masses at fixed $T$ are lower at higher $z$ as expected from the
virial theorm. ICM mass fractions show a mild $T$ dependence.
Clusters with $T \lta 3$~keV contain ICM mass fractions depressed
by modest amounts ($\lta 25\%$) below the cosmic mean baryon fraction
$\Omega_b/\Omega_m$; hot clusters subject to preheating 
remain good tracers of the cosmic mix of clustered mass
components at redshifts $z \lta 1$. 

\end{abstract}


\section{Introduction}


   The hot, \xray emitting plasma known as the intracluster medium
   (ICM) represents a part of the baryonic matter of the universe
   that is not associated with individual galaxies but remains trapped 
   in the deeper gravitational potential of galaxy clusters.  
   Models of cluster formation in which the intergalactic gas simply falls
   into the dark matter-dominated gravitational well (so-called infall
   models) fail to reproduce all the structural properties of the local
   cluster population (\eg Evrard \& Henry 1991; Navarro, Frenk \&
   White 1995; Mohr \& Evrard 1997; Bryan \& Norman 1998).  There appears 
   to be additional physics driving ICM evolution.

   As early \textit{ROSAT} and
   \textit{Einstein} data emerged, several researchers proposed that
   the missing element is the existence of a high entropy intergalactic gas
   prior to a cluster's collapse (David \etal 1991; Evrard \& Henry
   1991; Kaiser 1991; White 1991).  The entropy floor preduced by
   the preheated gas breaks
   the self-similarity between the dark matter and ICM on
   different mass scales.  This occurs because the equivalent
   thermal energy of the entropy floor corresponds to a larger fraction 
   of the virial temperature in lower mass clusters.

   A physical scenario consistent with this picture is one in
   which the bulk of star formation in a proto-cluster region
   occurs early in its formation history.  Heating from
   supernova-driven galactic winds and AGN activity associated
   with the rapid star formation ultimately comes to exceed the
   cooling rate of ambient material, resulting in a period of net
   local heating.  In the low-density limit, for which subsequent
   cooling is unimportant, this feedback defines an initial
   adiabat $S_i$ that is conserved in the absence of shocks.
   Later shock heating can raise the adiabat above this ``entropy
   floor,'' but the entropy cannot decrease below this level
   unless cooling becomes important.


Observational lines of evidence provide at least partial support for 
such a picture.  Measurements of ICM metal--abundances and their gradients
(\eg Loewenstein \& Mushotzky 1996) are likely explained by
feedback from star formation.  Most of the stars in cluster
elliptical galaxies are of a nearly uniform old age (Bower, Lucey
\& Ellis 1992; Kuntschner 2000, but see Trager etal 2000) and 
late-type galaxies within clusters are typically gas-poor and forming
stars at a reduced rate compared to field galaxies 
(Cayette \etal 1994, Chamaraux, Balkowski \& G\'{e}rard 1980).  
Even in MS1054, a
very distant cluster at $z \se 0.83$, much of the visible galaxy mass
is contained in red, early-type galaxies (Van Dokkum \etal
1999).  
An active period of star formation in at least some regions of the $z
\ssim 3$ universe is inferred from the ultraluminous sources detected
by SCUBA (Barger, Cowie \& Richards 2000), 
and the population of Lyman-break
galaxies, the distant equivalent of the local normal population, is
forming stars at a rate substantially higher than today (\seg,
Steidel \etal 1999).

In addition to winds driven by star formation, the jets of radio
galaxies present another potential source of ICM heating.  \chandra\ 
observations of low emission measure bubbles in
the ICM, aligned with the radio halos of central cluster sources
(McNamara \etal 2000), 
implicate radio jets as another source of additional ICM entropy.  The
fact that the radio galaxies appear more frequently in clusters at
higher redshift (Owen 2000) suggests that heating from this
population might be more important in the past.  

Of course, these and
other observations also indicate that some late star formation and
feedback does occur in cluster environments and this fact ultimately
limits the accuracy and applicability of the approach used here.
Also, the statistics of apparent gaps in the Ly-$\alpha$ forest
of QSO spectra can be used to place limits on the volume fraction
and epoch of heating to temperatures well above $10^4$ K (Theuns,
Mo \& Schaye 2000).  

A principal benefit of a model where clusters form through the
   infall of preheated gas is that the physical treatment
   of ICM evolution through gravitationally-driven shock heating is
   familiar territory for cosmological gas dynamic codes.  Comparing
   the gravitational evolution of a single \xray cluster derived from
   twelve nearly-independent codes, Frenk \etal (1999) find agreement
   in structural properties at characteristic levels of $\ssim 5\%$
   for dark matter and $\ssim 10\%$ for the ICM.

   The price for
   operating in a regime where the numerical accuracy has been
   calibrated is an approximate physical treatment.  Neglect of
   radiative cooling is a particular concern, but 
   the effect of radiative cooling and subsequent star formation 
   on a cluster's thermal history is a 
   complex problem.  Observations indicate that ICM cooling occurs
   frequently in clusters, but the phenomenon is restricted to a core 
   region containing $\ssim 1\%$ of the collapsed (virial) cluster
   mass.  However, recent spectroscopic studies which resolve
   cooling flows (David \etal 2000, Tamura \etal 2000) indicate
   that the phenomenon is confined to an even smaller volume than
   previously thought.

   At early times, cooling is an important ingredient of the
   galaxy formation process (Rees \& Ostriker 1977; White \& Rees
   1978).  So important, in fact, that sources of
   heating must be introduced to stabilize its effects on small scales
   (White \& Frenk 1991; Cole 1991; Blanchard, Valls-Gabaud \& Mamon
   1992).  Attempts to model the 
   full problem of the galaxy formation process within clusters are
   limited by poor knowledge of the correct parameterizations for star
   formation and feedback at arbitrary epochs.  Semi-analytic methods
   (Kauffman, White \& Guiderdoni 1993;  Baugh, Cole \& Frenk 1996;
   Wu, Fabian \& Nulsen 1998, 1999; Somerville \& Primack 1999) 
   are best able to explore the large parameter space
   associated with these processes, but the lack 
   of a direct solution for the spatial distribution of material
   limits the ability of this approach to predict detailed structural
   observables.  Bryan (2000) has presented a model which shows
   that the elimination of low--entropy gas reproduces the
   luminosity--temperature scaling relation.


\setcounter{footnote}{0}

   An initial entropy 
   excess prevents gas from falling into the dark
   matter--dominated potential well to the extent it would in the
   purely gravitational infall model, thus reducing the central gas
   density.  Core entropy\footnote{In this paper, an equivalent entropy $S$ is
   expressed in units $\kevcmsq$.  The usual thermodynamic
   definition of entropy is $s~=~c_V~ln(P/\rho^\gamma)$, where
   $\gamma~=~5/3$ for a monotonic gas (appropriate for a
   fully ionized plasma), relates to the entropy of this paper by 
   $S~=~e^{(s/c_V)}/R$, assuming an ideal gas equation of state
   $P~=~R~\rho~T$.} is
   related to temperature and density by $S_{core} \sim
   T/{\rho_{core}}^{2/3}$.  If the temperature is set by the
   collisionless dark matter through the virial theorem $T
   \spropto M_{200}^{2/3}$ ($M_{200}$ is the mass enclosed in a
   sphere defined as containing a density contrast of $200$ with respect to the
   critical density), then the core density of initially preheated material
   will satisfy
\begin{equation}
	\rho_{core} ~ \lta ~ M_{200} \ / \ S_i^{3/2} .
\label{rhocore}
\end{equation}
   Higher levels of preheating will produce lower core densities and,
   for a given preheating amplitude $S_i$, low mass clusters will
   have less dense cores than high mass clusters.  This differential
   effect results in a steepening of the slopes of the relations
   between density-dependent cluster observables (isophotal--size,
   luminosity and ICM mass) and temperature.


   Previous simulations have found preheating to produce more
   realistic clusters in terms of the luminosity--temperature relation
   (\eg Narvarro, Frenk \& White 1995; Pierre, Bryan \& Gastaud 1999).
   Metzler (1995) demonstrated a similar result using simulations that
   explicitly included the feedback of energy and mass from cluster
   galaxies.  We seek here to systematically investigate preheating's
   effect on the size--temperature, luminosity--temperature, and ICM
   mass--temperature relations.  Although not entirely independent
   observables, the functional dependence on the ICM
   density varies for these measures.  Each probes a
   differently-weighted moment of the radial density profile.  There
   is no guarantee, therefore, that a single preheating level $S_i$
   will provide a simultaneous match to all three relations.


   We use a total of 84 moderate resolution numerical experiments
   consisting of twelve initial configurations run at six different
   levels of preheating plus a set evolved without preheating for
   comparison.  Since we lack a detailed mechanism for the preheating,
   all simulations are initiated with the gas held at a fixed,
   elevated temperature that corresponds to adiabatically evolved
   temperatures of $\sim 10^6 K$ at $z \ssimeq 3$.


   In Section \ref{sec:tech}, the
   simulations are detailed.  The effect of preheating on the
   major cluster relations are presented in Section
   \ref{sec:scal}.  In Section \ref{sec:predict}, the range of
   entropies allowed by observation are determined and compared to
   other preheating studies.  The effects on evolution, the virial
   mass--temperature relation, and gas fractions are explored in
   Section \ref{sec:consequence}.

\section{Techniques}
\label{sec:tech}
\subsection{The Simulations}

    The simulations are run using the Lagrangian code P3MSPH (Evrard
    1988).  The parameters of the models we employ are similar to that
    used for the P3MSPH contribution to the Santa Barbara cluster
    comparison study (Frenk \etal 1999).  Initial conditions for our
    clusters are produced using the same multi-step procedure used in
    the Frenk \etal study.  First, purely gravitational N-body runs
    are used to identify an ensemble of clusters for resimulation with
    gas dynamics.  The final resimulation uses a combination of
    collisional and collisionless particles with a full gas dynamic
    treatment for a (Lagrangian) subset of the volume that comprises
    the cluster.  An intermediate, low resolution N-body model is used
    to identify the Lagrangian region to be treated with gas dynamics.

    The clusters are formed in a cold dark matter cosmology (Peebles
    1982; Blumenthal \etal 1984) dominated by a non-zero cosmological
    constant, ${\Lambda}$CDM.  The models assume a flat ($\Omega = 1$)
    geometry with the following parameters: $\Omega_{m} = 0.3$,
    $\Omega_{\Lambda} = 0.7$, $\Omega_{b} = 0.03$, $\sig8 = 1.0$,
    $\Gamma = 0.24$ and $h = 0.8$.  The Hubble constant is defined as
    $100~h \kmsmpc$; and \sig8 is the power spectrum normalization on
    $8h^{-1}$ Mpc scales.  Clusters form from initial density
    perturbations which are Gaussian random fields consistent with a
    CDM transfer function specified by the shape parameter,
    $\Gamma \equiv {\Omega}h$ (Bond \& Efstathiou 1984).  Initial
    conditions are constructed using Zel'dovich's formulation
    (\eg Efstathiou \etal 1985).


    The simulations begin with two $128^3$ N-body runs which represent
    cubic regions 366 Mpc on a side from a CDM power spectrum.
    Clusters that form in this region have masses as large as $10^{15}
    M_\odot$ containing ${\sim}10^3$ particles.  We identify six
    clusters to resimulate from the final configurations of the two
    volumes: the two most massive clusters, a random pair less massive
    by a factor 3 from the mean of the first pair, and another random
    pair reduced in mass by a further factor 3.  This results in a
    resimulated sample comprised of twelve clusters covering roughly a
    decade in mass.

    The density field in cubic regions centered on the chosen
    clusters' initial states are extracted from the original
    simulation and placed onto a higher resolution $64^3$ grid.  The
    size of the regions ranges from 50-100 Mpc and scales as the cube
    root of the mass enclosed within the turn around radius of the 
    resimulated cluster.  High frequency modes
    of the density field not sampled by the original simulation are
    then added, up to the new limiting Nyquist frequency.  From this
    set of initial conditions the cluster is evolved in a purely
    N-body simulation of $32^3$ particles, using alternate sampling of
    the $64^3$ density field.  Particles in this simulation that lie
    within a density contrast of 5 with respect to the background are
    used to define a Lagrangian mask.  For the final gas dynamic
    resimulation, masked locations in the $32^3$ subsampled field are
    expanded by a factor of two in resolution, generating an effective
    $64^3$ resolution within the non-linear regions of the cluster.
    The high-resolution inner regions contain between 20,424 and
    26,064 dark matter and gas particles in equal numbers, where
    the gas particles are placed to trace the dark matter.  The inner
    regions are surrounded by 29,510 to 30,215 low-resolution dark
    matter particles that are each eight times the mass of a combined
    high-resolution gas and dark matter pair.  This treatment allows
    the inclusion of both tidal effects from the surrounding large
    scale structure and the gas dynamics of the virial region in an
    economical way.  The tidal particle contamination inside $r_{200}$
    at $z=0$ is $\leq 0.1\%$ by mass.  In addition, the cluster resimulations
    have mass resolutions which are similar fractions of the cluster 
    virial masses despite spatial resolutions that
    range by a factor two, from 125-250 kpc.


    Preheating is achieved by specifying an initial hot temperature
    for the baryons.  The initial epoch is defined by a linear growth
    factor of 16 to the present day, resulting in an initial redshift
    $z_i \se 20.82$ for {$\Lambda$}CDM.  The initial temperature of the 
    gas in the six preheating models ranges from $1.5-9.5{\times}10^7$ K 
    compared to $10^4$ K in infall models.  Since we expect that
    this heating actually occurs at $z \sim 3$, this method of
    preheating is valid only if a small amount of gas has
    collapsed at the true epoch of preheating.  In these 12
    clusters 4 -- 19\% of the gas particles which eventually
    collapse have done so at $z \sim 3$.

    Table~1 lists the temperatures and corresponding initial entropies
    $S_i$ for the seven models.  The latter are derived using the cosmic
    mean baryon density at the initial epoch.  Since the simulations
    have density perturbations at the outset, the entropy floor is not
    perfectly flat.  Also listed in Table~1 are values of the gas
    temperature at $z \se 3$ that would place gas at the background density
    on the same initial adiabat.  Energetically, at this epoch,
    the temperatures correspond to values between 0.07 and 0.4
    keV per particle.  

    Table~\ref{SIMtab} lists basic properties of the complete set of
    84 clusters at $z \se 0$.  Groups are numbered by mass beginning with the
    highest mass group from each of the two initial N-body models
    (labeled ``a'' and ``b'').  Individual runs are labeled by
    combining their model and group numbers with the initial entropy
    level (\eg, a190S3).

	\footnotesize \begin{center}
	{\sc TABLE 1\\ Model Properties}
	\vskip4pt 
	\begin{tabular}{lccc} 
	\hline \hline 
	Model & $S_i$ & $T_{adiab} (z=3.0)$ & $T_i (z=20.82)$ \\
	 & [keV cm$^2$] & [K] & [K] \\ 
	\hline 
	S6 & 335.4 & $5.3 \times 10^6$ & $9.5 \times 10^7$\\ 
	S5 & 221.8 & $3.4 \times 10^6$ & $6.0 \times 10^7$\\
	S4 & 141.2 & $2.2 \times 10^6$ & $4.0 \times 10^7$\\ 
	S3 & 105.9 & $1.7 \times 10^6$ & $3.0 \times 10^7$\\ 
	S2 & 88.3 & $1.4 \times 10^6$ & $2.5 \times 10^7$\\
	S1 & 53.0 & $8.4 \times 10^5$ & $1.5 \times 10^7$\\ 
	S0 & 0.035 & $5.6 \times 10^2$ & $1.0 \times 10^4$\\ 
	\hline
	\end{tabular} 
	\vskip2pt 
	\end{center} 
	\setcounter{table}{1}
	\normalsize

\subsection{Data Processing}

	Mathiesen \& Evrard (2001; hereafter ME01) find that there are
	significant differences (up to 20\%) between the temperatures
	that theorists generally report (\eg mass-weighted
	temperature) and their observed counterparts (which are
	based on plasma models).  The values reported here build on the work of
	ME01 in order to be as close to observed values as possible.
	Luminosities and isophotal sizes are determined using the
	XSPEC \ttfamily mekal \normalfont spectral emission model with
	0.3 solar metallicity on a bandpass from 0.5 to 2.0 keV where
	the clusters are imaged at z=0.06.  Unless otherwise stated,
	cluster temperatures used below are spectral values determined
	by the best fit of the cluster emission for a fixed
	photon count within $r_{500}$ to an
	isothermal \ttfamily mekal \normalfont spectrum in a 0.5 to
	10.0 keV bandpass.

\section{The Scaling Relations}
\label{sec:scal}
	Clusters exhibit a number of scaling relations that lend
	insight into their physical nature.  A cluster's \xray
	isophotal size is tightly correlated with its
	emission weighted mean temperature; the so-called size--temperature 
	relation (\sst; Mohr \& Evrard 1997; Mohr et al. 2000; hereafter M00).  
	The relationship between a cluster's luminosity and its
	temperature (\slxt) has higher scatter, (David et al. 1993), reflecting
	its dependence on the ICM density
        distribution (the luminosity) and the total binding mass (temperature). 
        This relation is important in determining the temperature function
	of a flux-limited sample of clusters.  Because variations in ICM mass fraction
	are rather modest (Mohr, Mathiesen \& Evrard 1999; hereafter MME99),
	the ICM mass correlates with cluster temperature leading to an \mt
	relation.  In addition to being useful in constraining
	cosmological parameters, these relations can be used to
	constrain models of the interaction between galaxies and the ICM.

\subsection{Scaling Formalism}
\label{sec:formalism}

	The self-similar model of Kaiser (1986) is a useful starting
	point in developing the expected behavior of the scaling
	relations to be considered in this section.  The model assumes
	a smooth, spherically symmetric distribution of gas and total
	mass about its center.  We follow notation similar to that of 
	Arnaud \& Evrard (1999).  Let $M_{\dc}$ be the total mass
	contained in a sphere (of radius $r_{\dc}$ about the center)
	that encompasses a mean density ${\dc}\rho_c$, where
	$\rho_{c}(z){\equiv}3H(z)^2/8{\pi}G$ is the critical density
	and $H(z)$ the Hubble parameter of the universe at epoch $z$.
	We write the ICM density ${\rho}_{ICM}(r)$ in terms of the
	natural radial variable $y{\equiv}r/r_{\dc}$
\begin{equation}
	{\rho}_{ICM}(yr_{\dc})~\equiv~f_{ICM} \, {\dc} \, \rho_c \,
	g(y)
\label{gasdeneqn}
\end{equation}
	where explicit use of the ICM mass fraction within a density
	contrast $\dc$
\begin{equation}
	f_{ICM}~\equiv~\frac{M_{ICM}(<r_{\dc})}{M_{\dc}}
\label{gasnorm}
\end{equation}
	sets the normalization of the structure function
	$3\int_{0}^{1}dy~y^2~g(y)=1$.

	A population of clusters that is strictly self--similar will
	have constant $f_{ICM}$ and a single, specific function
	$g(y)$.  Given the random and fully three-dimensional nature
	of cluster formation dynamics, strict self-similarity is an
	unrealistic expectation (Jing \& Suto 1998; Thomas \etal
	2000).  A more reasonable expectation is that clusters are a
	{\sl regular\/} population whose gas fraction values and
	structure functions will exhibit deviations depending, for
	example, on dynamical history.  Searching for trends in the
	population with temperature, it is useful to employ mean
	gas fractions and structure functions
\begin{equation}
\ficmhat(T) ~\equiv~\langle f_{ICM} \rangle_T ,
\label{ficmhat}
\end{equation}
\begin{equation}
\ghat(y|T) ~\equiv~\langle g(y) \rangle_T
\label{ficmhat}
\end{equation}
	where $\langle ~ \rangle_T$ denotes an ensemble average over
	clusters with temperature $T$.

	Assume that the virial theorem holds, so that the mass $M_\dc$
	and temperature are linked by
\begin{equation}
	M_\dc~=~ C_m \, T^{\alpha_m}
\label{VT}
\end{equation}
	where the exponent $\alpha_m$ is $3/2$ if spectral temperature
	is an unbiased measure of the mass-weighted value.  Depending
	on the applied band pass, simulations display a bias that
	increases $\alpha_m$ by $\ssim 10-20\%$ (Mathiesen \& Evrard
	2001; and see \S5.2 below).

	Finally, approximating cluster radial temperature profiles as
	isothermal, we can write the temperature scaling relations for
	$M_{ICM}$, $L_X$ and $R_I$ that are expected from a regular
	cluster population described by mean gas fractions
	$\ficmhat(T)$ and structure functions $\ghat(y|T)$.  The ICM
	mass will scale as
\begin{equation}
	M_{ICM}~=~C_m \ficmhat(T)~T^{\alpha_m}
\label{MICMreln}
\end{equation}
	and the bolometric \xray luminosity as
\begin{equation}
	L_X~=~C_X \, \ficmhat(T)^2 \, Q_L(T) \, \tilde{\Lambda}(T) \,
	T^{\alpha_m}.
\label{Lreln}
\end{equation}
	where $\tilde{\Lambda}(T)$ is a dimensionless emissivity (see
	Appendix) and $Q_L(T) \sequiv 3 \int_0^1 dy \, y^2 \,
	\ghat^2(y|T)$.  The size at fixed isophote requires an
	additional assumption about the projected radial profile of
	the emission.  We assume a standard $\beta$-model for which
	$I(R) \spropto R^{1-6\beta}$ outside the core.  For this case,
	the expected scaling of isophotal size is
\begin{equation}
	R_I~=~C_R \, [ \ficmhat^2(T) \, Q_I(T) \, \tilde{\Lambda}(T)
	\, T^{2\alpha_m\beta} ]^{1/(6\beta-1)}
\label{RIreln}
\end{equation}
	with $Q_I(T) \sequiv \int_o^{\sqrt{1-\xi^2}} d\eta \,
	h^2(\sqrt{\xi^2+\eta^2})$ and $0.1 \gta \xi^2 \lta 1$.  For
	the typical value $\beta \se 2/3$ (Jones \& Forman 1984; MME99),
	this becomes
\begin{equation}
	R_I~=~C_R \, \ficmhat^{2/3}(T) \, Q_I^{1/3}(T) \,
	\tilde{\Lambda}^{1/3}(T) \, T^{4\alpha_m/9} .
\label{RIreln2}
\end{equation}

	For self-similar clusters and mass-weighted temperatures,
	familiar scalings $M_{ICM} \spropto T^{3/2}$ and $L_X \spropto
	T^2$ emerge for the ICM mass and luminosity.  The isophotal
	size in this case scales as $R_I \spropto T^{2/3}$.
	Equations~(\ref{MICMreln}), (\ref{Lreln}) and (\ref{RIreln})
	demonstrate that deviations of observed slopes from these
	expectations can arise from a number of sources.
	Temperature-dependent mean gas fractions and structure
	functions may be largely responsible, but metallicity trends
	that affect $\tilde{\Lambda}(T)$ and a virial slope $\alpha_m
	\ne 3/2$ will also tilt these relations.

\subsection{The \mt Relation}


	Two-component $\beta$-model analysis of 45 clusters in the
   	\xray flux-limited Edge sample leads to an observed \mt
   	relation of the form (MME99)
\begin{equation}
	log(M_{ICM}) = (1.98{\pm}0.18)log(T_6) + (13.42{\pm}0.03) -
\frac{5}{2}log(h)
\label{MTfit}
\end{equation}
      	where $M_{ICM}$ is the ICM mass within a density contrast of
	500 in units of $h^{-1}M_\odot$ and $T_6 \se T/6$keV.  The
	slope $1.98\pm0.18$ is significantly different from the value
	of 3/2 expected from a strictly self-similar population.

\begin{figure*}
\begin{minipage}{160mm}
\epsfxsize=17.0cm \epsfysize=8.5cm

\hbox{\epsfbox{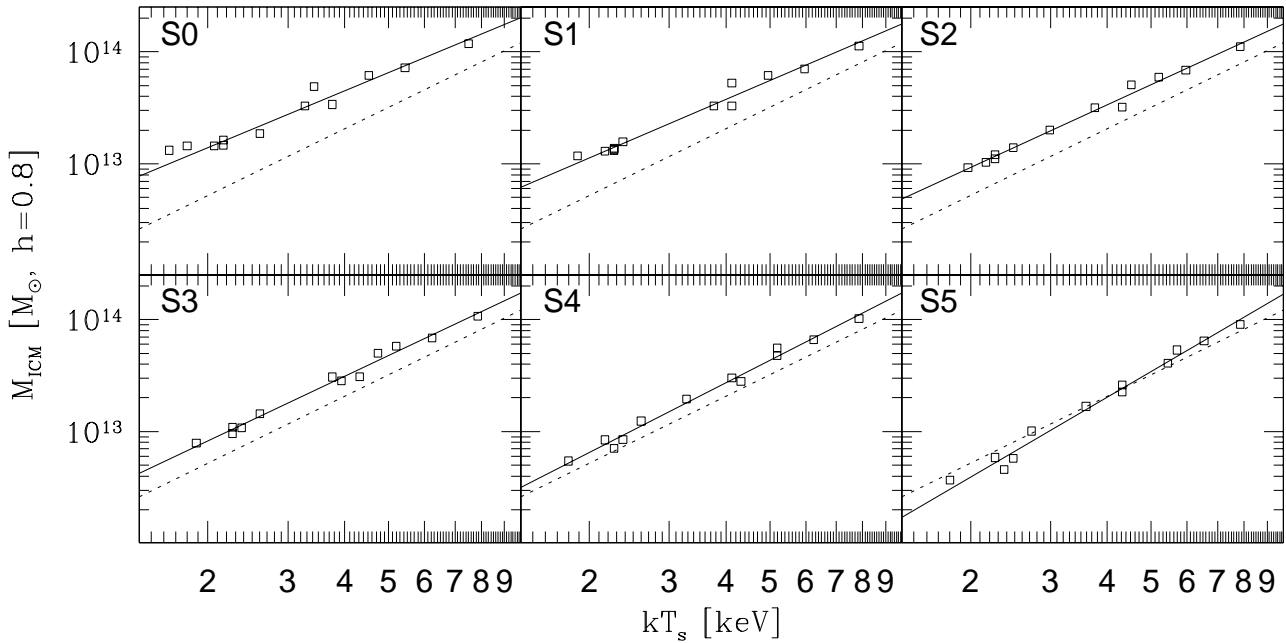} }

\caption{ The ICM mass--temperature relation at $\Delta_c \se 500$ for
12 clusters run at 6 initial temperatures.  The solid line is
the best fit to clusters with kT$_s > 2$ keV.  Dotted lines in each panel
show the MME99 observational result. }
\label{MTplot}
\end{minipage}
\end{figure*}

	Figure~\ref{MTplot} displays the \mt relations at $\Delta_c
	\se 500$ for the simulated cluster sets at zero redshift.  In
	this and subsequent figures of this section, the temperature
	is the spectral fit to emission within \rfiveh.  This radius
	is chosen because it is observationally accessible, but our
	results are fairly insensitive to this choice.  Slopes
	differ at the level of $0.6-6\%$ between
	$r_{500}$ and $r_{200}$.  The dotted line in each panel is the MME99
	observational result scaled to $h \se 0.8$.  Increased levels
	of preheating lead to steeper slopes in the \mt relation; the
	six preheated models sweep right through the observed
	slope (note: S6 is not displayed to conserve space).
	Fits for the model sets are listed in Table~\ref{OBStab}.
	These fits are made only for clusters with $kT_s > 2.0$
	keV, to reflect the range over which the MME99 fit was made.

	Lower mass clusters have a larger core relative to their
	virial radius with respect to high mass clusters.  This
	ratio of core radius to virial radius increases with
	increasing initial entropy.  Consequently, the fraction of
	gas on adiabats higher than the default self-similar case also
	increases.  This has the effect of lowering the gas mass
	fraction within the virial region; implying a decreased \micm.

	This argument can also be cast in terms of the extropy
	floor discussed earlier.  The entropy floor of the cluster
	effectively `plugs', in the sense of making Jeans stable, some
	amount of core material.  Later infalling matter encounters
	this material at progressively larger radii as $S_i$ is
	increased, leading to spillover at larger radii.  Lower mass
	clusters are affected by this to a larger degree because their
	virial adiabat is lower than that of rich clusters,
	eq.~(\ref{rhocore}).  The fit for each set of models is
	listed in Table~\ref{OBStab}.  These fits are made only for
	clusters with $kT_s > 2.0$ keV, to reflect the range over
	which the MME99 fit was made.

	Models S1--S4 have slopes in statistical agreement with the
	observed value.  The zero--points are modestly higher than
	observed, which may be an indication that the global
	baryon fraction of 0.1 used in the simulations is $\ssim
	20\%$ too high.  Care must be taken in making this
	comparison because the observed and simulated values
	scale differently with Hubble's constant ($h^{-5/2}$ and
	$h^{-1}$, respectively).  Re-scaling to $h\se 0.7$ would
	raise the observed values $22\%$ higher than the
	simulated measures, essentially eliminating the offset
	for model S3, for example.  Since other factors like ICM
	clumpiness due to mergers (Mathiesen, Evrard \& Mohr 1999) or
	a multiphase medium (Gunn \& Thomas 1996; Nagai, Sulkanen \&
	Evrard 2000) can affect the ICM mass at similar ($10-20\%$)
	levels, attempting to pin down the zero-point of this
	relation to a precision better than this remains problematic. 
	
\begin{figure*}
\begin{minipage}{160mm}
\epsfxsize=17.0cm \epsfysize=8.5cm

\hbox{\epsfbox{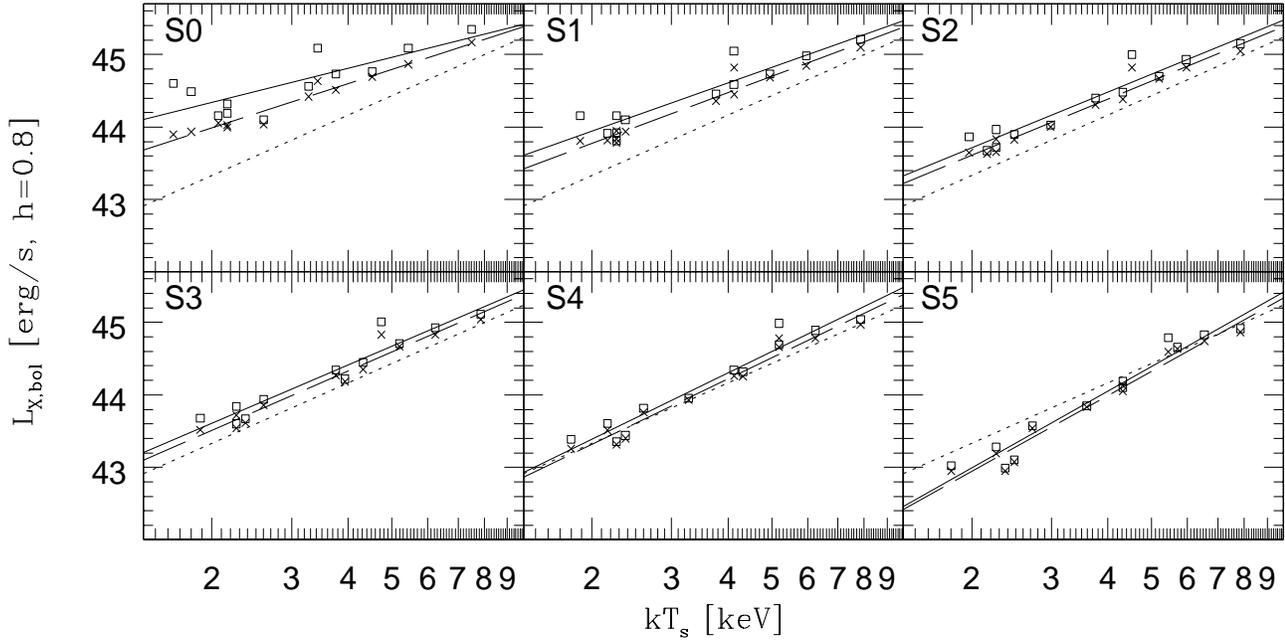} }

\caption{ The final bolometric \xray luminosity \vs spectral
temperature for the 12 clusters evolved from 6 initial temperatures.
Temperatures are estimated within $r_{500}$, luminosities are shown
within $r_{200}$ (open squares) or within that radius but with the
core emission (within $0.13 r_{200}$) removed (crosses).  Solid and
dashed lines are fits to these respective data.  The observational
result for $h \se 0.8$ is superimposed with a dotted line in each
panel.  }
\label{LxTplot}
\end{minipage}
\end{figure*}

\begin{figure*}
\begin{minipage}{160mm}
\epsfxsize=17.0cm \epsfysize=8.5cm

\hbox{\epsfbox{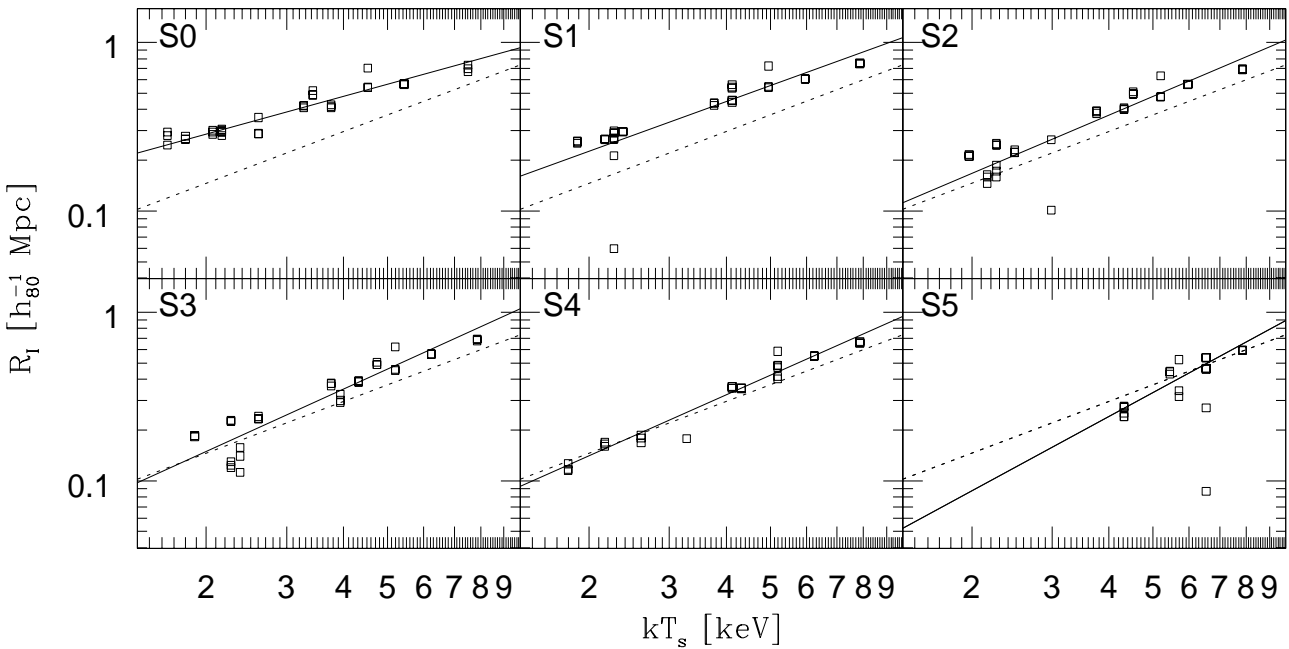} }

\caption{ The size--temperature relations for the ensemble are shown
for an isophote level $I = 3.0 \times 10^{-14}$
< ergs/s/cm$^2$/arcmin$^2$ in the 0.5 - 2.0 keV band.  The solid line is
the best fit to clusters with kT$_s > 2$ keV.  The observed relation
is represented by the dotted line.}
\label{STplot}
\end{minipage}
\end{figure*}

\setcounter{table}{2}
\begin{table*}
\begin{minipage}{160mm}
\centering
\caption{Fits$^a$ to the \mt, \lxt \& \st Relations at $z=0$ ($h=0.8$)}
\begin{tabular}{lcccccccc}	        
\hline\hline
&
\multicolumn{2}{c}{ $\mt$}&
\multicolumn{2}{c}{ $\lxt~(core-included)$} &
\multicolumn{2}{c}{ $\lxt~(core-extracted)$} &
\multicolumn{2}{c}{ $\st$}\\ \hline
Model & $m$ & $b$ & $m$ & $b$ & $m$ & $b$ & $m$ & $b$ \\ 
\hline
S6 &  2.67(0.13) & 13.58(0.03) & 3.96(0.26) & 44.36(0.07) & 3.95(0.24)
& 44.32(0.07) & - & - \\
S5 &  2.37(0.10) & 13.72(0.03) & 3.52(0.25) & 44.67(0.07) & 3.47(0.21)
& 44.60(0.06) & 1.47(0.04) & -0.36(0.01) \\
S4 &  2.07(0.08) & 13.80(0.02) & 3.03(0.24) & 44.84(0.07) & 3.00(0.18)
& 44.75(0.05) & 1.20(0.07) & -0.28(0.01)\\
S3 &  1.92(0.08) & 13.83(0.02) & 2.67(0.25) & 44.89(0.07) & 2.71(0.19)
& 44.81(0.06) & 1.25(0.14) & -0.24(0.02)\\
S2 &  1.86(0.07) & 13.85(0.02) & 2.55(0.24) & 44.93(0.07) & 2.57(0.17)
& 44.84(0.05) & 1.15(0.10) & -0.23(0.02)\\
S1 &  1.73(0.09) & 13.88(0.03) & 2.20(0.28) & 45.00(0.09) &
2.32(0.18) & 44.89(0.06) & 0.98(0.17) & -0.18(0.02)\\
S0 & 1.69(0.12) & 13.95(0.04) & 1.56(0.38) & 45.09(0.14) & 2.02(0.13)
& 44.96(0.05) & 0.74(0.05) & -0.19(0.02)\\
Observed  & 1.98(0.18) & 13.66(0.03) &  2.76(0.15) & 44.67(0.03) &
& & 1.02(0.11) & -0.35(0.02) \\
\hline
\end{tabular}
\label{OBStab}
\centerline{$^a$ Using the form $\log(X) = m \log(T_s/6 \kev) + b$.}
\end{minipage}
\end{table*}

\vskip20pt

\subsection{The $L_X - T$ Relation}

	The \lxt relation has received considerable attention from
	observers (\eg Edge \& Stewart 1991; David \etal 1993) and
	computational model builders (Metzler 1995; Navarro, Frenk \&
	White 1995; Allen \& Fabian 1998; Eke, Navarro \& Frenk 1998)
	over the last decade.  An important contribution was made by
	Fabian \etal (1994; \cf Allen \& Fabian 1998), who showed
	that excess core emission associated with cooling flows
	is the primary source of the large scatter observed in
	the \lxt relation.  The scatter is significantly reduced
	in analyses that either excise the core cooling flow
	regions (Markevitch 1998) or examine samples of clusters
	defined to possess weak cooling cores (Arnaud \& Evrard 1999).

	Since our simulations do not include the effects of cooling,
	we will compare them to the results of the studies that avoid
	cooling cores.  Analyzing 26 clusters with accurate
	temperatures and inferred cooling rates $\dot{M} \le 100 \msol
	\yr^{-1}$, Arnaud \& Evrard (1999) find the correlation
	between bolometric luminosity and temperature $L \propto T^q$
	with $q=2.88 \pm 0.15$.  Markevitch (1998) removes photons
	from the central $100 \hinv \kpc$ when calculating both the
	temperature and luminosity and this produces $q=2.64 \pm
	0.27$.  We use an unweighted average of these two
	observational results, leading to an observed relation
	with intermediate slope
\begin{equation}
log(L_X) = (2.76 \pm 0.15)~ log(T_6) + (44.48 \pm 0.03) - 2 log(h).
\label{LxTobs}
\end{equation}

	The bolometric \lxt relations for the model sets at a redshift
	of zero are displayed in Figure~\ref{LxTplot}.
	Power--law fits are listed in Table~\ref{OBStab}.
	Infall models (S0) exhibit a slope $1.44 \pm 0.26$ that
	is marginally inconsistent with the $L \propto T^2$
	self-similar expectation (see Section \ref{sec:formalism}).
	The slope is biased by the two lowest temperature systems;
	each lies a factor of roughly three above a fit performed
	under their exclusion.  Examination of the pair's density
	profiles revealed that their core internal gas density profiles are
	steeper compared to the rest of the sample.  A possible
	explanation of these enhancements is that they are transient
	effects of fortuitously observed mergers (Roettiger \etal
	1996).  Exclusion of core emission, where the core is defined
	as a circular area of radius $0.13 r_{200}$ (Neumann \& Arnaud
	1999), yields a best fit \lxt slope $2.04 \pm 0.16$, a value
	consistent with the analytic scaling.  In
	Figure~\ref{LxTplot}, crosses show the core-extracted \lxt
	relation and the dashed line shows the best fit.

	Low-mass clusters, with lower virial entropy, feel the effects
	of preheating to a greater degree than high-mass clusters, 
	resulting in a
	steepening \lxt relation as $S_i$ is increased.  
	The slopes of models S2--S4 are 
	in agreement with observations.  In the preheated
	models, no significant change in slope is seen by neglecting
	the core (see Table~\ref{OBStab}).  The change in behavior
	compared to the S0 models is expected when one considers the
	fact that the core density is bounded from above by the
	imposed initial entropy, eq.~(\ref{rhocore}).

	The ability to match the \lxt slope is not unique to the
	form of preheating assumed here.  Metzler's (1995)
	models employing continuous energy feedback produce a slope of
	$2.96{\pm}0.05$.  The \lxt relation is also explained by
	semi-analytic models such as the shock model of Cavaliere,
	Menzi \& Tozzi (1999) and models based on energy input and
	hydrostatic arguments 
	(Wu, Fabian \& Nulsen 1998, 1999; Bower \etal 2000).  Bryan
        (2000) matches the \lxt relation with a model that removes low
        entropy ICM gas and assumes it to be trapped within galaxies.

\subsection{The \st Relation}

   	The isophotal size $R_I$ of a nearby cluster is tightly
  	correlated to its temperature in a power law relation with
  	slope near unity (Mohr \etal 2000, hereafter M00)
\begin{equation}
	log~ R_I = (1.02 \pm 0.11)~ log(T_6) - (0.44 \pm 0.01) -
	log(h)
\label{STfit}
\end{equation}
	with $R_I$ in units of Mpc.  Mohr \& Evrard (1997) showed that
	the observed slope is steeper than the range $0.61 \le m \le
	0.81$ found in sets of purely gravitational simulations of
	clusters in four different CDM cosmologies.  The isophotal
	sizes from an ensemble of models experiencing continuous
	feedback showed a slope $0.99$,
	consistent with observations.
   
	Figure~\ref{STplot} illustrates the effect of preheating on
	the \st relation at zero redshift.  The size $R_I$ is
	determined at an isophote level $I = 3.0 \times 10^{-14}$
	ergs/s/cm$^2$/arcmin$^2$ in the 0.5 - 2.0 keV band.  In
	each panel, an individual cluster contributes three data
	points, one for each orthogonal projection.   The
	conversion from count rate to physical flux units is made
	by using PROS, assuming a Raymond--Smith spectrum with
	the published mean temperature $T_X$ and $\frac{1}{3}$ solar abundances
	(Mushotzky \& Loewenstein 1997).  The angular diameter
	distance is used to convert angular size into physical
	size.  $I$ is also corrected for cosmological dimming, $(1+z)^4$.   
	Table~\ref{OBStab}
	lists the best fit power law parameters for the simulations.
	Only clusters with kT$_s > 2$ keV are used in the fit, so as
	to match the range of the observed dataset used by M00.
	The fit is the average of 10,000 sets of 11 randomly
	drawn simulation images.  Assuming that 2 out of 3 projections can
	be treated as independent, each set excludes a random projection.
	
	Preheating leads to
	steeper slopes; the six models produce a range of slopes that
	encompass the observed value. Model S1 is in best agreement
	with observation, but models S2 through S4 are also consistent.

\section{Constraining the Preheating Level}
\label{sec:predict}

Preheating succeeds in reproducing the slopes of the three observed
	scaling relations considered in the preceding section.
	Figure~\ref{mplot} summarizes the results by displaying the
	slope fit to each of the preheated sets along with the
	observational constraints for the \mt, \lxt and \st relations.
	Simultaneous agreement in all three slopes can be satisfied
	within a range of initial entropy values near $\ssim 100
	\kevcmsq$. 

	We use a chi--squared measure based on the slopes of the three
	relations $\Delta\chi^2 \se \sum_{i=1}^3 \,
	(m_{sim,i}-m_{obs,i})^2/(\sigma_{sim,i}^2+\sigma_{obs,i}^2)-\chi^2_{min}$
	for each set of models in order to constrain the range of allowed
	preheating.  Figure~\ref{xplot} shows the chi--squared for
	each relation individually and their sum.  Horizontal lines
	indicate the 68.3\%, 90\% and 99\% confidence limits for one
	degree of freedom.  From this analysis, the predicted level
	of preheating required to reproduce observation is between
	55--150 keV cm$^2$ at 90\% confidence.


	This range overlaps values derived independently by
	Lloyd-Davies, Ponman \& Cannon (2000).  In an analysis of
	\xray emission profiles of groups and clusters, they inferred
	a trend of decreasing central density with decreasing
	temperature, consistent with the constraint of a minimum
	central entropy in the range 34-82 keV cm$^2$ with 90\%

\myputfigure{figure4.epsi}{2.8}{0.45}{-10}{-0} \figcaption{The \mt,
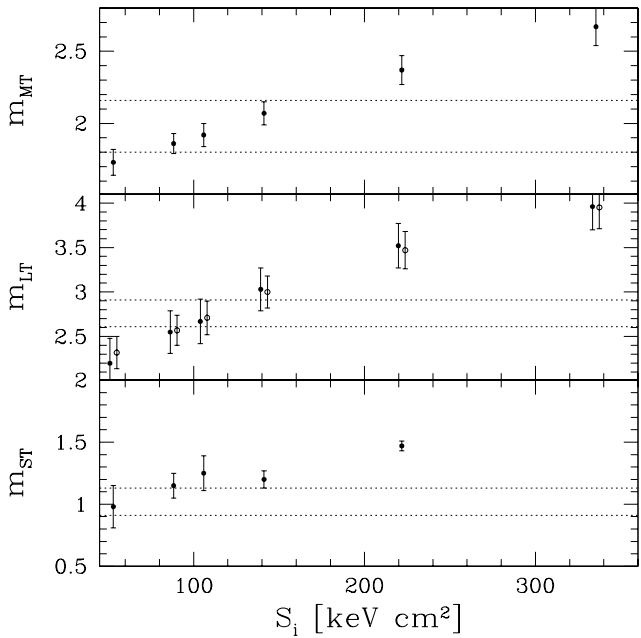
\lxt, and \st slopes as a function of initial entropy.  The allowed
slopes based on observation are indicated by a pair of dotted lines.
All errors are $1\sigma$.  The open symbols in the \lxt panel
represent the core-extracted fits (the two symbols have
been offset about the $S_i$ value for clarity). 
\label{mplot}}
\vskip10pt

\noindent
	certainty.  Pierre, Bryan, and Gastaud (1999) produced a
	suite of simulated clusters at a preheated entropy of 141
	keV cm$^2$ and achieve results that are in agreement with the
	observed \lxt relation (see also Pen 1999).  Balogh, Babul \&
	Patton claim success with a model in which the gas at the
	turnaround density is heated to $1.8 \times 10^6 \K$ at $z \se
	1$, translating (for $\Omega_m \se 0.3$, $h\se 0.8$) 
	into an equivalent entropy value $170 \kevcmsq$. 

	As previously mentioned, preheating is not unique in its ability
	to solve the \lxt scaling relation problem.  Semi--analytical
	models (Wu, Fabian and Nulson 1999a,b; Bower \etal 2000; Tozzi
	\& Norman 2000) typically add energy into the ICM in a more
	gradual fashion.  Since the evolutionary history of the ICM
	would differ in different scenarios, the behavior of the distant
	cluster population should decide between ICM histories based
	on rapid or more gradual heating.  We turn next to 
	predictions for ICM evolution at $z \se 0.5$ and $z \se 1$. 

\section{Evolutionary Predictions}
\label{sec:consequence}

Preheating affects the evolution of the ICM in a characteristic way.
Observations of the distant cluster population can therefore be used
to falsify the model or define the limits of its applicability.
Figures~\ref{SBimg} through~\ref{SZimg} give a representation
of the effect of preheating on cluster a1.  
The projected \xray emissivity,
thermal SZ signal, and emission-weighted temperature at redshifts 0,
$0.5$, 1 and 2 are shown for the same cluster under three treatments:
no heating (model S0), a successful level of preheating (S3) and an
excessively heated case (S6).  The surface-brightness and temperature
maps are made using the \ttfamily mekal \normalfont emission model and
the images are sized to be $2r_{200}$ on a side. The excessive model
S6 shows the power of preheating to deplete the central ICM density of
a cluster and severely limit its \xray visibility at high redshift.
The impact on the preferred model S3 is less dramatic.  In

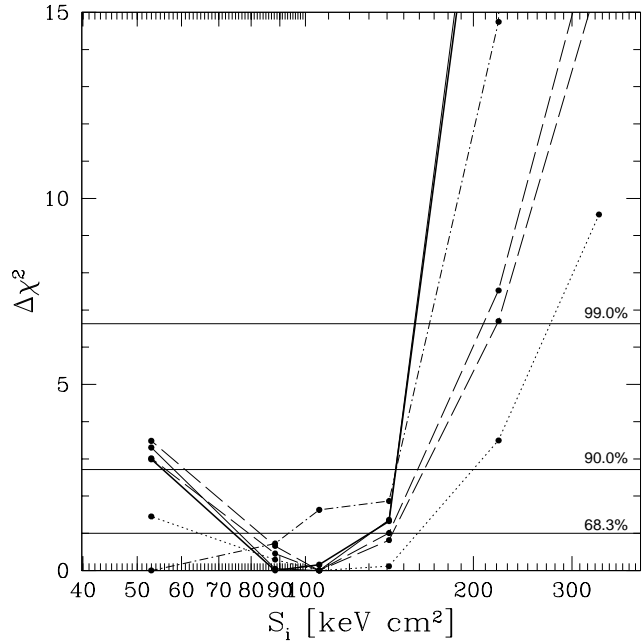
\myputfigure{figure5.epsi}{2.8}{0.45}{-10}{-0} \figcaption{The
$\Delta\chi^2$ measure for the scaling relation slopes is shown as a
function of initial entropy.  The dotted, heavy dashed, dot--dashed lines
correspond to the \mt, \lxt and \st relations, respectively.  The
solid line indicates the $\chi^2$ when considering all three
relations.  The light dashed and solid lines correspond to the use of the
core-extracted \lxt fits.
\label{xplot}}
\vskip10pt 

\noindent this
section, we present the redshift evolution of the locally successful
set of S3 models only.  This model is chosen because it lies
mid--range in the allowed initial entropies and because it is the
best fit for two of the three relations.

\subsection{Evolution of Scaling Relations}

Figures~\ref{MTevofig}, \ref{LxTevofig} and \ref{STevofig} show
evolution in the \smt, \lxt and \st relations, respectively, for model
S3 at redshifts 0, 0.5 and 1. Table~\ref{evotab} provides fits to
power law relations.  The ICM mass and luminosity are measured within
$r_{200}$ at each redshift and the physical isophotal size is
determined at a fixed intrinsic isophote level of $I = 3.0 \times
10^{-14}$ergs/s/cm$^2$/arcmin$^2$ in the 0.5 - 2.0 keV band.

The simulations predict little evolution in the \st and \lxt relations
to $z$ = 0.5.  However, at $z$ = 1.0, both relations become $\ssim
30\%$ steeper.  Current observations of high redshift clusters appear
consistent with these predictions.  Mohr \etal (2000) find no
significant evolution in their intermediate redshift study
($z~\approx$ 0.2--0.5) of CNOC clusters imaged by ROSAT.  There is no
significant evolution in the \lxt relation to redshifts $\ssim 0.5$
(Mushotzky \& Scharf 1997; Fairley \etal 2000).  Higher quality data
from the coming \chandra\ and XMM/Newton archives will tighten the
constraints on the degree of allowed evolution.  

	Figure~\ref{MTevofig} illustrates that the \mt relation evolves
	toward lower ICM masses at higher redshift for fixed
	temperature clusters, as expected in the canonical evolution model.  
	As with the \st and \lxt relations, no
	significant evolution in slope is expected at redshift 0.5
	while a moderate steepening to $M_{ICM} \spropto T_s^{2.4 \pm
	0.2}$ is seen at $z \se 1$.  Because ICM masses within
	$r_{200}$ require very deep imaging, less observational
	information is available for this re-lation at high redshifts.
	The expectations of model S3 ap-

\setcounter{figure}{8} 
\myputfigure{figure9.epsi}{2.8}{0.45}{-10}{0} \figcaption
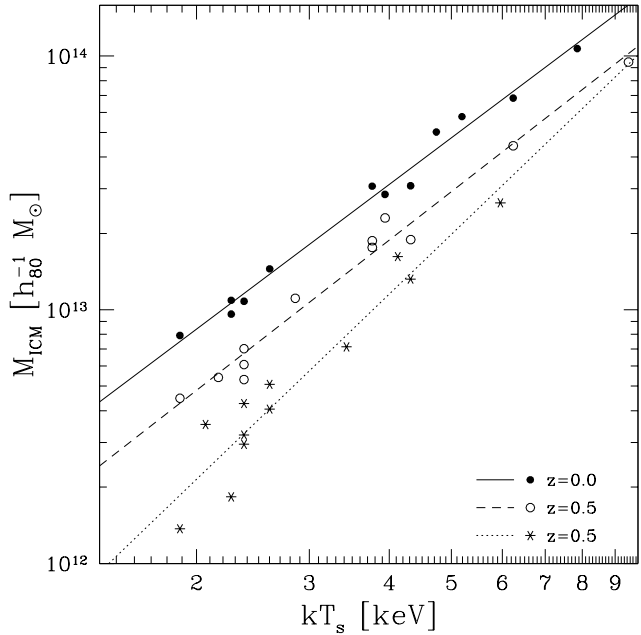
{ The ICM mass--temperature relation for the clusters of preheated
model S3 at three redshifts.
\label{MTevofig}}
\vskip10pt

\noindent ppear consistent with
	observations (\eg, Matsumoto \etal 2000).  

\subsection{Virial Mass \& ICM Mass Fractions}

	The fact that the \mt relation of set S3 is steeper than the
	canonical $3/2$ expectation arises from two separate factors.  
	One is the relation between total mass
	and temperature and the other is the ICM mass fraction within
	$r_{200}$.  

	Non-preheated models follow a relation between
	total mass $M_{\Delta_c}$ and the mass-weighted temperature
	$T_m$ that is consistent with expectations from the virial
	theorem $H(z) M_{\Delta_c} \spropto T_m^{3/2}$
	(e.g. Bryan \& Norman 1998; M00; ME01).  When a $2-10$ keV
	spectral temperature is employed in place of the mass-weighted
	value, ME01 find that the relation steepens slightly to 
	$H(z) M_{\Delta_c} \spropto T_m^{1.6}$.  Our
	non-preheated set of models (S0) is consistent with this
	steepening.  

	For the preheated set S3, 
	Figure~\ref{virplot} presents the relation between total
	mass within $r_{200}$ versus spectral
	temperature at $z$ = 0, 0.5 and 1.  Power-law fits to the data, 
	listed in Table~5, indicate that slopes of $1.66 \pm 0.12$
	and $1.73 \pm 0.11$ at $z=0$ and $0.5$, respectively, are
	consistent with the spectral relation of the non-preheated
	models.  At $z \se 1$, the slope steepens to $1.91 \pm 0.11$.  
	Note that the intercept at $T_s \se 6 \kev$ remains stable,
	varying by only a few percent over the redshift range probed.

Observational determination of this relation can be attempted using
	weak lensing masses.  Current data are too noisy
	to discriminate between a slope of $1.5$ and $1.6$ (Hjorth, Oukbir
	\& Van Kampen 1999).  Weak lensing masses are likely biased by the
	structures in which the clusters are embedded (Metzler et al. 1999);  
	In addition to lensing masses, estimates based on
	galaxy kinematics (Girardi \etal 1998; Horner, Mushotzky
	\& Scharf 1999) or the hydrostatic assumption with
	measured tem-

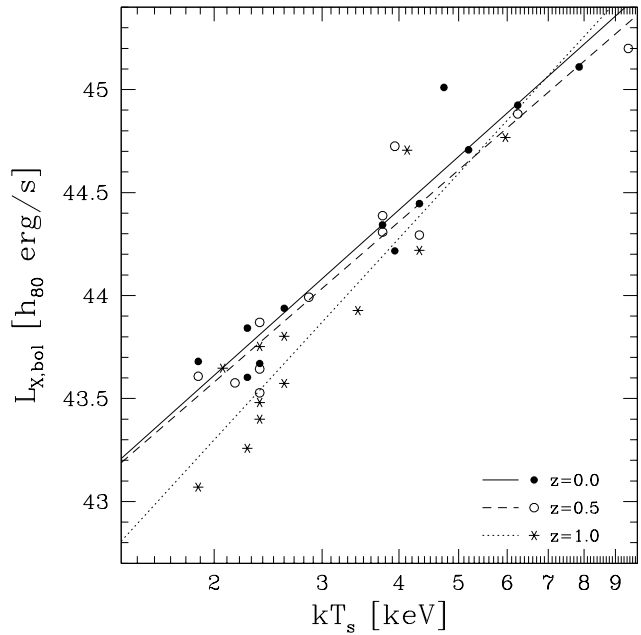
\myputfigure{figure10.epsi}{2.8}{0.45}{-10}{0} \figcaption{ The
luminosity--temperature evolution for the clusters of model S3.
\label{LxTevofig}}
\vskip10pt

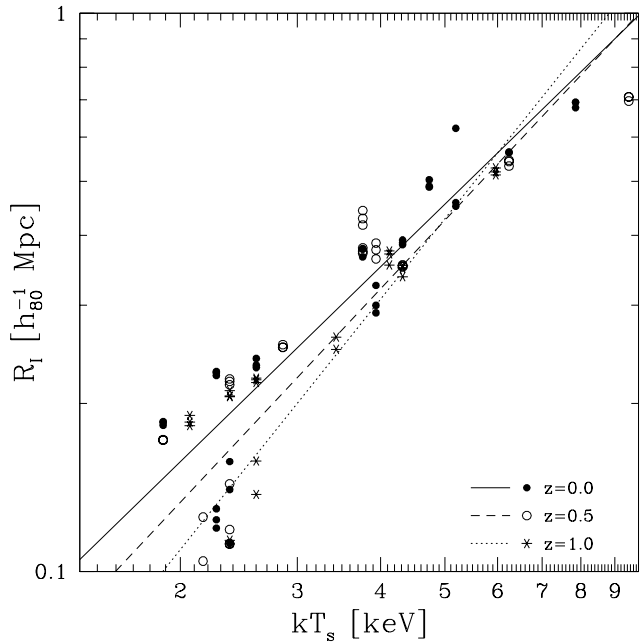
\myputfigure{figure11.epsi}{2.8}{0.45}{-10}{0} \figcaption{The
size--temperature relation for model S3 at $z$ = 0, $0.5$ and 1.
\label{STevofig}}
\vskip10pt

\noindent
perature profiles are possible.  Nevalainen,
	Markevitch \& Forman (2000) follow the latter approach
	for a sample of nine galaxies, groups and clusters and
	find a relation with slope $1.79 \pm 0.14$, steeper by
	$3.6\sigma$ than the canonical $1.5$ expectation but more
	in line with the relation from set S3.  Finoguenov,
	Reiprich \& Boehringer (2000) analyze a larger sample and
	find a slope of $1.78^{+0.10}_{-0.09}$, consistent with
	the Nevalainen et al (2000) results and inconsistent with
	self similar expectation at more than 3$\sigma$.

The ICM mass--temperature relations are steeper than that for the
total mass at all redshifts for set S3, because the ICM mass
fraction decreases at lower cluster temperatures.
Figure~\ref{gasplot} shows that the baryon fraction within
$r_{200}$ decreases from $9\%$ at high temperatures to about
$6\%$ at 2~keV.  Note that all values are consistently lower than
the cosmic baryon fraction of 10\%, an effect generally seen even
in non-preheated simulations (MME99; Frenk \etal 1999).  The
fall-off at low temperatures is evidence of preheating's stronger
impact on shallower gravitational potentials.

        \begin{table*}
	\begin{minipage}{160mm}
	\centering
\caption{Fits$^a$ to the \st, \lxt, \mt relations for preheating set
S3 ($h=0.8$)} 
	\begin{tabular}{lcccccc}
	\hline
	\hline
	& \multicolumn{2}{c}{\mt} 
	& \multicolumn{2}{c}{\lxt} 
	& \multicolumn{2}{c}{\st} \\
	\hline
	$z$ & m & b & m & b & m & b \\
	\hline
	0.0 & 1.90(0.07) & 13.83(0.02) & 2.67(0.25) & 44.89(0.07) & 1.16(0.12) & -0.25(0.02) \\
	0.5 & 1.97(0.09) & 13.62(0.03) & 2.59(0.25) & 44.81(0.08)
	& 1.27(0.19) & -0.27(0.05) \\
	1.0 &  2.43(0.23) & 13.49(0.08) & 3.25(0.43) & 44.85(0.15) & 1.49(0.19) & -0.25(0.03)\\
	\hline
	\label{evotab}
	\end{tabular} 
\vskip -8truept
\centerline{$^a$ Using the form $\log(X) = m \log(T_s/6 \kev) + b$.}
	\end{minipage}
	\end{table*}

	\noindent 
  Because cluster
potentials tend to deepen with time, clusters evolve along this
relationship toward values of higher ICM mass fraction.  

	Preheating and its effects on the ICM mass fraction and
	the \mt relation is of direct relevance to the observability of 
	high redshift clusters based on their \xray emissivity or the
	thermal Sunyaev-Zel'dovich effect.  The latter is
	particularly well suited for high redshift cluster
	detection.  Using non-preheated simulations which with
	constant ICM mass fraction, Holder \etal (2000) show that
	proposed interferometric arrays would be capable of
	detecting clusters above a total mass limit of $10^{14} \hinv \msol$ at
	essentially any redshift.  Haiman, Mohr \& Holder (2000)
	emphasize the power of such surveys to constrain
	cosmological parameters $\Omega_M$, $\Omega_\Lambda$ and
	the equation of the state parameter $w$ of the dark
	energy component.  An analysis of the effects of
	preheating on these yields and possible cosmological
	constraints is underway (Mohr \etal in prep.).


\section{Conclusions}


	We have used a set of 84 moderate resolution gas dynamic
	simulations of cluster evolution to systematically investigate 
	the effects of 
	preheating on the local \st, \lxt and \mt relations.  We
	use a plasma emission model to estimate luminosities and
	determine spectral temperatures from fits to the overall cluster
	emission.  
	We find that preheated models with initial entropy in the range $55-150
	\kevcmsq$ reproduce the slopes of the observed scaling relations.  

        There is agreement
        between the observations of Lloyd-Davies \etal (2000) and
        these simulations in the allowed range of initial 
        entropy.  Although the range is currently generous, the new
        generation of \xray satellites should provide information 
        to further constrain the allowed range or potentially falsify
        the model.  Better understanding of the ICM entropy aids
        in the deduction of its history and places restrictions on the
        mechanisms that may be responsible for preheating.

       Examining evolutionary effects in a set with 
       $S_i \se 106 \kevcmsq$, the models predict modest steepening
       in the three scaling relations at $z = 1.0$.  The predictions
       are currently in agreement with 

\footnotesize
\begin{center}
{\sc Table 5\\ Fits$^a$ to $h(z) M_{200}$ \vs spectral temperature for set S3 }
\vskip 4pt
\begin{tabular}{ccc}
\hline \hline z & m & b \\ \hline 
0.0 & 1.66(0.12) & 14.98(0.04) \\
0.5 & 1.73(0.11) & 14.95(0.04) \\ 
1.0 & 1.91(0.11) & 14.96(0.04) \\
\hline
\end{tabular}
\centerline{$^a$ $\log(h(z)M_{200})~=~m \log(T_s/6 \kev) + b$.}
\end{center} 
\setcounter{table}{5} \normalsize \centerline{}

\myputfigure{figure12.epsi}{2.8}{0.45}{-10}{0} \figcaption{
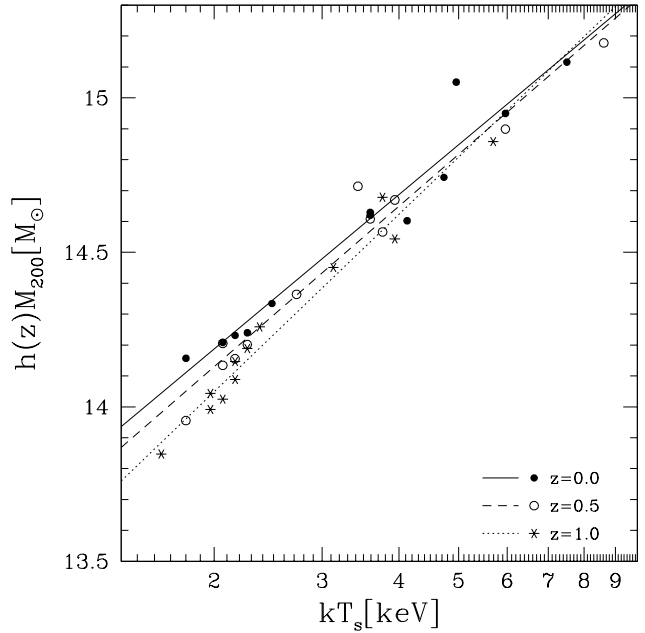
Redshift-corrected total mass within $r_{200}$ \vs spectral
temperature for clusters in set S3.
\label{virplot}}
\vskip10pt

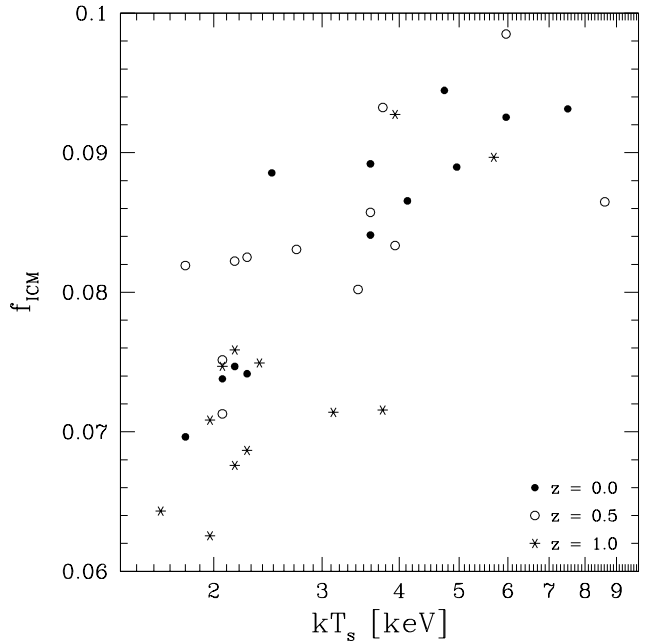
\myputfigure{figure13.epsi}{2.8}{0.45}{-10}{0} \figcaption{ICM
mass as a fraction of total mass \vs spectral temperature inside
$r_{200}$ for clusters of model S3 at $z$ = 0.0, 0.5 and 1.0.  The
global value is $\Omega_b/\Omega_m = 0.1$.
\label{gasplot}}
\vskip10pt

\noindent
        the observed lack of evolution in the \lxt relation to $z
        \ssim 0.5$.  Deeper observations (\eg, Stanford \etal 2000) 
        from Chandra and XMM-Newton should help reveal the
	epoch where the scalings depart from the simulation's expectations.


\acknowledgments

The simulations were done, in part, using the computing facilities at
the University of Michigan's Center for Parallel Computing.  
Our data processing was improved using \ttfamily mekal\normalfont.  We would
like to thank Ben Mathiesen and Martin Sulkanen for their help with
the spectral models.  We would also like the thank Greg Bryan and
an anonymous referee for many helpful comments on an earlier
version of this manuscript. This work was supported by
NASA through grant NAG5-7108 and NSF through grant AST-9803199.
JJM is supported by Chandra Fellowship grant PF8-1003, awarded
through the Chandra Science Center.  The Chandra Science Center
is operated by the Smithsonian Astrophysical Observatory for NASA
under contract NAS8-39073.

\begin{appendix}

	Let $M_{\dc}$ be the total mass contained in a sphere (of
	radius $r_{\dc}$ about the center) that encompasses a mean
	density ${\dc}\rho_c$, where
	$\rho_{c}(z){\equiv}3H(z)^2/8{\pi}G$ is the critical density
	and $H(z)$ the Hubble parameter of the universe at epoch $z$.
	Write the ICM density ${\rho}_{ICM}(r)$ in terms of the
	natural radial variable $y{\equiv}r/r_{\dc}$, structure
	function $g(y)$, and the ICM gas fraction within $r_{\dc}$, as
	in eqn's~(\ref{gasdeneqn}) and (\ref{gasnorm}).

	Clusters have no well-defined edge, but their hydrostatic 
	regions are reasonably well bounded by $r_{200}$.  We normalize the
	total cluster luminosity to be that interior to this radius 
\begin{equation}
	L_X~=~4\pi \, \int_{0}^{r_{200}}~dr\, r^2~n_e(r)~n_H(r)~{\Lambda}(T(r))
\label{Leqn}
\end{equation}
	where ${\Lambda}(T(r))$ is an appropriately normalized
	emissivity dependent only on temperature and
	$n_e=\rho_{ICM}/\mu_e m_p$ and $n_H=\rho_{ICM}/\mu_H m_p$ are
	the number densities of free electrons and protons.  
	Assuming that the
	ICM is isothermal, we can define a dimensionless emissivity
	$\tilde{\Lambda}(T) \sequiv {\Lambda}(T)/{\Lambda}(10 \kev)$ by
	arbitrarily normalizing to emission at $T \se 10 ]\kev$.  
	Using equations~(\ref{gasdeneqn}) and (\ref{gasnorm}), the luminosity
	can be rewritten as 
\begin{equation}
	L_X~\equiv~ C_X ~{f_{ICM}}^2 ~Q_L ~\tilde{\Lambda}(T) ~ T^{\alpha_m}
\label{LQeqn}
\end{equation}
	where $C_X$ carries dimension and depends only on fundamental
	constants and  $Q_L=3
	    \int_0^1~dy\,y^2\,g^2(y)$.  
	Note that $Q_L$ is equivalent to the clumping factor 
	$<\rho_{ICM}^2>/<\rho_{ICM}>^2$, where the
	angle brackets denote the volume average over the cluster atmosphere.
	It is thus a structure factor which depends solely on the gas
	density shape and characterizes the concentration of the gas
	distribution.

	Allowing expectations for $f_{ICM}$ and $Q_L$ to vary with
	cluster temperature produces eq.~(\ref{Lreln}).  
	To see the origin of the luminosity's canonical
	infall model temperature dependence, a set of additional
	assumptions are traditionally made: (i) pure bremsstrahlung
	emission ($\Lambda(T)~\propto~T^{1/2}$), (ii) virial
	equilibrium ($M_{200}~\propto~T^{3/2}$), (iii) structurely
	identical clusters ($Q_L(T)~=~C_1$) and (iv) constant gas
	fraction ($f_{ICM}~=~C_2$).  This leads to the expected
	scaling relation between luminosity and temperature:
	$L_X~\propto~T^2$.

	The scaling for the \st relation is outlined in Mohr \etal
	(2000).  Assuming no emission beyond $r_{200}$, then the 
	surface brightness at $r_{200}$ is formally zero.  Consider
	the surface brightness at radius $\xi r_{200}$ where $\xi$ is
	near, but below, unity 
\begin{equation}
	{\Sigma}(\xi r_{200})~=~\frac{1}{2\pi(1+z)^4}\int_{-r_{200}\sqrt{1-\xi^2}}^{r_{200}\sqrt{1-\xi^2}}dl~n_e(r)~n_H(r)\Lambda(T)
\label{SIGeqn}
\end{equation}
	where $r=\sqrt{(\xi r_{200})^2+l^2}$.
        Rewriting using dimensionless variables as in the luminosity
        case above, we find 
\begin{equation}
	{\Sigma}(\xi r_{200})~=~ C_I~{f_{ICM}}^2 ~Q_I ~\tilde{\Lambda}(T)~ T^{\alpha_m/3}
\label{sigleqn}
\end{equation}
	where $C_I$ is a constant and $Q_I \se \int_0^{\sqrt{1-\xi^2}}
	d\eta \, g^2(\sqrt{\eta^2 + \xi^2})$.  

        Assume that the $\beta$-- model (Cavaliere \& Fusco--Femiano 1976)
	describes the surface brightness profile
\begin{equation}
	\Sigma(R)~=~\Sigma_0~\left(1~+~\left(\frac{R}{R_c}\right)^2\right)^{-3{\beta}+\frac{1}{2}}
\label{Beqn}\end{equation}
	Outside the core, the surface brightness scales as a power law
	in radius $\Sigma(R)~\propto~R^{-6\beta+1}$.  
	In this regime, the surface brightness at an isophotal distance,
	$\Sigma(R_I)$, can be related to the surface brightness at
	radius $\xi r_{200}$ 
\begin{equation}
	R_I~=~\xi r_{200}~\left(\frac{{\Sigma}(\xi r_{200})}{{\Sigma}(R_I)}\right)^{1/(6\beta-1)}
\label{RIeqn}
\end{equation}
	This adds a constraint on the shape parameter $Q_I \spropto
	\xi^{1-6\beta}$.  
	Using eq.~(\ref{sigleqn}) and introducing temperature dependence
	in $f_{ICM}$ and $Q_I$ then leads to the result 
	eq.~(\ref{RIreln}) of \S3.1.  
	For a typical observed value of $\beta~=~2/3$ along with
	canonical assumptions for the other parameters, the
	predicted size temperature relation is $R_I~\propto~T^{2/3}$.

\end{appendix}

\setcounter{table}{1}
\begin{deluxetable}{lccccccccr}
\tablewidth{0pt} \tablecaption{Cluster properties ($\dc=200, z=0$)}
\tablehead{ \colhead{Model} & \colhead{ L\tablenotemark{a} } &
\colhead{ $M_{tot}$\tablenotemark{b} }& \colhead
{ $M_{ICM}$\tablenotemark{c} }& \colhead{ $\Upsilon$ } & \colhead
{ $\sigma_{DM}$\tablenotemark{d} } & \colhead{ $T_m$\tablenotemark{e}
}& \colhead{ $T_s$\tablenotemark{f} }& \colhead
{ $L_{X,bol}$\tablenotemark{g} }& \colhead{ $R_I$\tablenotemark{h} }}
\startdata
 a1S6 & 72.53 & 8.48 & 7.02 & 0.828 & -85 & 5.72 & 6.24&
5.65 & 0.3719\nl
 a2S6 & 72.53 & 3.01 & 1.88 & 0.625 & -303& 3.06 &
2.99& 0.408 & -\nl
 a31S6 & 53.34 & 3.90 & 2.54 & 0.651 & -18 & 3.69 &
3.94& 0.937 & -\nl
 a34S6 & 42.86 & 1.76 & 0.872& 0.495 & -91 &
2.23 & 2.48& 0.164& -\nl
 a190S6& 37.89 & 1.37 & 0.395& 0.288 & -117&
1.89 & 2.07& 0.0377& -\nl
 a260S6& 31.04 & 0.896& 0.318& 0.355 & -30 &
1.52 & 1.80& 0.0498& -\nl
 b1S6 & 82.43 & 10.4 & 8.80 & 0.846 & -311&
6.41 & 5.96& 4.18& -\nl
 b2S6 & 86.26 & 11.7 & 10.4 & 0.889 & 84
& 7.81 & 8.23& 8.13& -\nl
 b31S6& 51.56 & 5.31 & 3.31 & 0.623 & -208&
4.43 & 4.96& 2.64& -\nl
 b34S6 & 48.23 & 3.52 & 1.82 & 0.517 &
-157& 3.35 & 3.59& 0.476& -\nl
 b233S6 & 40.91 & 1.48 & 0.485& 0.328 &
-53 & 2.06 & 2.27& 0.0558& -\nl
 b255S6 & 32.99 & 1.30 & 0.493& 0.379 &
-67 & 1.85 & 2.16& 0.0692& -\nl
 \hline
 a1S5 & 72.53 & 8.96 & 7.89 & 0.881 & -51 & 5.91 & 6.24& 8.39 &
0.4308\nl
 a2S5 & 72.53 & 3.58 & 2.66
& 0.743 & -331& 3.10 & 3.13& 0.889 & -\nl
 a31S5 & 53.34 & 4.17 &
3.35 & 0.803 & -37 & 3.65 & 3.94& 1.93 & 0.2215\nl
 a34S5 & 42.86 &
1.95 & 1.41 & 0.723 & -80 & 2.27 & 2.48& 0.470 & -\nl
 a190S5 &37.89 & 1.43 & 0.655& 0.458 & -61 & 1.85 & 2.16& 0.122& -\nl
 a260S5 &31.04 & 0.952& 0.497& 0.522 & -91 & 1.48 & 1.72& 0.131& -\nl
b1S5& 82.43& 11.4 & 10.2 & 0.895 & -316& 6.23 & 5.44& 5.69 &
0.2519\nl
 b2S5 & 86.26 & 12.1 & 11.4 & 0.942 & 69 & 7.65 & 7.50& 10.2
& 0.4759\nl
 b31S5 & 51.56& 5.38 & 4.34 & 0.807 & -22 & 4.69 & 5.19&
7.71 & 0.3442\nl
 b34S5& 48.23 & 3.74 & 2.73 & 0.730 & -62 & 3.58 &
3.94& 1.58 & 0.1922\nl
 b233S5& 40.91 & 1.58 & 0.775& 0.491 & -87 &
1.99 & 2.27& 0.160& -\nl
 b255S5 & 32.99 & 1.41 & 0.753& 0.534 & -52 &
1.85 & 2.16& 0.239& -\nl
 \hline
 a1S4 & 72.53 & 8.88 & 8.08 &
0.910 & -74 & 5.79 & 5.96& 9.70 & 0.4395\nl
 a2S4 & 72.53 & 3.70 & 2.96
& 0.800 & -343& 2.93 & 2.99& 1.12& -\nl
 a31S4 & 53.34 & 4.34 &
3.75 & 0.864 & -63 & 3.57 & 3.94& 2.73 & 0.2908\nl
 a34S4 & 42.86 &
2.06 & 1.76 & 0.854 & -50 & 2.36 & 2.48& 0.822& 0.1435\nl
 a190S4 &37.89 & 1.50 & 0.944& 0.629 & -64 & 1.79 & 2.07& 0.283& -\nl
a260S4 & 31.04 & 1.23 & 0.760& 0.618 & -188& 1.46 & 1.72& 0.305&
0.0922\nl
 b1S4 & 82.43 & 11.5 & 10.2 & 0.887 & -309& 5.96 & 5.19&
6.12& 0.3221\nl
 b2S4 & 86.26 & 13.0 & 12.1 & 0.931 & 130 & 7.76 &
7.50& 15.4& 0.5218\nl
 b31S4 & 51.56& 5.46 & 4.93 & 0.903 & 30 & 4.75 &
4.96& 12.1& 0.3749\nl
 b34S4 & 48.23 & 3.98 & 3.29 & 0.827 & -17 & 3.70
& 3.94& 2.61 & 0.2854\nl
 b233S4 & 40.91 & 1.65 & 1.07 & 0.648 & -86 &
1.93 & 2.16& 0.345 & -\nl
 b255S4 & 32.99 & 1.62 & 1.05 & 0.648 &
-87 & 1.84 & 2.07& 0.507 & 0.1283\nl
 \hline
 a1S3 & 72.53 & 8.88 & 8.24
& 0.928 & -53 & 5.80 & 5.96& 10.5 & 0.4500\nl
 a2S3 & 72.53 & 2.51 &
3.50 & 0.839 & -216& 3.35 & 3.59& 2.06 & 0.2398\nl
 a31S3 &53.34& 2.72
& 3.80 & 0.892 & -40 & 3.41 & 3.59& 2.20 & 0.3031\nl
 a34S3 &42.86 &
1.37 & 1.91 & 0.884 & -71 & 3.40 & 2.48& 1.08& 0.1873\nl
 a190S3 &37.89 & 0.853 & 1.19 & 0.735 & -71 & 1.83 & 2.07& 0.501& 0.0990\nl
a260S3 & 31.04 & 0.716 & 1.00 & 0.694 & -174& 1.59 & 1.80& 0.598&
0.1485\nl
 b1S3 & 82.43 & 7.16 & 10.0 & 0.893 & -300& 5.81 & 4.96&
6.37 & 0.3608\nl
 b2S3 & 86.26 & 8.70 & 12.2 & 0.938 & 110 & 7.81 &
7.50& 16.1 & 0.5418\nl
 b31S3& 51.56 & 3.74 & 5.22 & 0.944 & 116 & 4.69
& 4.73& 12.8 & 0.3905\nl
 b34S3 & 48.23 & 2.48 & 3.46 & 0.865 & 41 &
3.76 & 4.12& 3.49 & 0.3139\nl
 b233S3 & 40.91 & 0.922 & 1.29 & 0.741 &
-42 & 1.98 & 2.27& 0.585& 0.1122\nl
 b255S3 & 32.99 & 0.910 & 1.27 &
0.747 & -13 & 1.94 & 2.16& 0.869& 0.1825\nl
 \hline
 a1S2 & 72.53 & 8.88
& 8.32 & 0.937 & -78 & 5.71 & 5.96& 10.7 & 0.4495\nl
 a2S2 & 72.53 &
3.75 & 3.18 & 0.848 & -369& 2.79 & 2.72& 1.34& 0.0808\nl
 a31S2 &53.34& 4.29 & 3.80 & 0.886 & -39 & 3.43 & 3.59& 3.15 &
0.3145\nl
 a34S2& 42.86 & 2.05 & 1.82 & 0.888 & -68 & 2.28 & 2.37& 0.991& 0.1788\nl
a190S2& 37.89 & 1.59 & 1.26 & 0.792 & -87 & 1.82 & 2.07& 0.593&
0.1315\nl
 a260S2 & 31.04 & 1.50 & 1.12 & 0.747 & -158& 1.65 & 1.88&
0.920& 0.1727\nl
 b1S2 & 82.43 & 11.1 & 9.84 & 0.886 & -293& 5.72 &
4.96& 6.26 & 0.3780\nl
 b2S2 & 86.26 & 13.0 & 12.2 & 0.938 & 104 &
7.79 & 7.50& 17.7 & 0.5484\nl
 b31S2 & 51.56& 5.62 & 5.30 & 0.943 & 108
& 4.61 & 4.52& 12.4& 0.3943\nl
 b34S2 & 48.23 & 4.09 & 3.55 & 0.868 &
28 & 3.77 & 4.12& 3.81 & 0.3283\nl
 b233S2 & 40.91& 1.72 & 1.53 & 0.890
& -83 & 1.90 & 2.16& 0.655& 0.1376\nl
 b255S2 & 32.99& 1.74 & 1.37 &
0.787 & 5 & 1.99 & 2.16& 1.16& 0.2009\nl
 \hline
 a1S1 & 72.53 & 8.82 & 8.24 & 0.934 & -47 & 5.73 & 5.69 & 12.1 & 0.4865\nl  
 a2S1 & 72.53 & 2.44 & 2.04 & 0.836 & -196& 2.21 & 2.27 & 0.811& 0.0478\nl
 a31S1 &53.34&  4.42 & 3.87 & 0.876 & -18 & 3.50 & 3.59 & 3.60 & 0.3485\nl
 a34S1& 42.86 & 2.10 & 1.94 & 0.924 & -41 & 2.24 & 2.37 & 1.57 & 0.2369\nl
a190S1& 37.89 & 1.71 & 1.49 & 0.871 & -53 & 1.90 & 2.16 & 1.03 & 0.2120\nl
a260S1 & 31.04 &1.59 & 1.36 & 0.855 & -152& 1.72 & 1.88 & 1.80 & 0.2082\nl
 b1S1 & 82.43 & 11.2 & 9.60 & 0.857 & -257& 5.63 & 4.96 & 6.76 & 0.4378\nl
 b2S1 & 86.26 & 13.4 & 12.5 & 0.933 & 92  & 7.80 & 7.50 & 20.3 & 0.5947\nl
 b31S1 & 51.56& 5.84 & 5.48 & 0.938 & 143 & 4.57 & 4.32 & 14.0 & 0.4281\nl
 b34S1 & 48.23 &3.98 & 3.52 & 0.884 & -81 & 3.64 & 4.12 & 4.84 & 0.3605\nl
 b233S1 & 40.91&1.74 & 1.52 & 0.874 & -7  & 1.96 & 2.16 & 1.05 & 0.2162\nl
 b255S1 & 32.99&1.78 & 1.54 & 0.865 & 2   & 2.01 & 2.16 & 1.81 & 0.2389\nl
\hline
 a1S0 & 72.53 & 8.96 &
8.32 & 0.929 & -72 & 5.55 & 5.19& 15.4 & 0.4547\nl
 a2S0 & 72.53 & 3.41
& 2.81 & 0.824 & -308& 2.52 & 2.48& 1.58& 0.2304\nl
 a31S0& 53.34 &
4.40 & 3.84 & 0.873 & -26 & 3.28 & 3.28& 4.56 & 0.3349\nl
 a34S0 &42.86 & 2.11 & 1.93 & 0.915 & -73 & 2.13 & 2.07& 2.59 & 0.2410\nl
a190S0 &37.89 & 1.86 & 1.66 & 0.892 & 5 & 1.95 & 2.07& 1.79&
0.2283\nl
 a260S0 &31.04 & 1.75 & 1.55 & 0.886 & -136& 1.53 & 1.64&
5.05 & 0.2358\nl
 b1S0 & 82.43 & 10.7 & 9.12 & 0.852 & -202& 5.44 &
4.32& 7.33 & 0.4339\nl
 b2S0 & 86.26 & 13.6 & 12.6 & 0.926 & -77 &
7.73 & 7.17& 28.0 & 0.5353\nl
 b31S0 &51.56& 5.95 & 5.41 & 0.909 & 147
& 4.19 & 3.42& 15.4 & 0.3923\nl
 b34S0 &48.23 & 4.06 & 3.58 & 0.882 &
-42 & 3.59 & 3.59& 6.78 & 0.3337\nl
 b233S0 &40.91& 1.80 & 1.59 &
0.883 & 18 & 1.94 & 2.07& 1.94& 0.2373\nl
 b255S0 &32.99 & 1.92 &
1.64 & 0.854 & 3 & 1.85 & 1.80& 3.84 & 0.2228\nl \enddata
\tablenotetext{a}{[ $h^{-1}$ Mpc ]} \tablenotetext{b}
{[ $h^{-1}~10^{14}M_\odot$ ]} \tablenotetext{c}
{[ $h^{-1}~10^{13}M_\odot$ ]} \tablenotetext{d}{[ km/s ]}
\tablenotetext{e}{[ K ]} \tablenotetext{f}{[ K ]} \tablenotetext{g}
{[ $h~10^{44}$ erg/s ]} \tablenotetext{h}{[ $h^{-1}$ Mpc ],
dashed entries result from no emission above the chosen isophote }
\label{SIMtab}
\end{deluxetable}

\clearpage

	\setcounter{figure}{5}
	\figcaption{\xray surface--brightness (SB) images for
	cluster a1 are shown for non--preheated (S0, left),
	moderately preheated (S3, middle) and excessively
	preheated (S6, right) cases at four redshifts (0.0, bottom;
	0.5; 1.0; 2.0, top).\label{SBimg}}

	\vskip10pt

	\figcaption{\xray temperature maps for
	cluster a1 are shown for non--preheated (S0, left),
	moderately preheated (S3, middle) and excessively
	preheated (S6, right) cases at four redshifts (0.0, bottom;
	0.5; 1.0; 2.0, top).\label{Timg}}

	\vskip10pt

	\figcaption{\sz (SZ) Compton Y images for
	cluster a1 are shown for non--preheated (S0, left),
	moderately preheated (S3, middle) and excessively
	preheated (S6, right) cases at four redshifts (0.0, bottom;
	0.5; 1.0; 2.0, top).\label{SZimg}}

	\vskip10pt


\begin{thebibliography}{}

\bibitem[ ] { } Allen S.W. \& Fabian A.C., 1998, \mnras, 297, L57.

\bibitem[ ] { } Arnaud M. \& Evrard A.E., 1999, \mnras, 305, 631

\bibitem[ ] { } Balogh M.L., Babul A., Patton D.R., 1999, \mnras, 307,
463

\bibitem[ ] { } Barger A.J., Cowie L.L. \& Richards E.A., 2000,
\apj, 119, 2092


\bibitem[ ] { } Baugh C.M., Cole S. \& Frenk C.S., 1996, \mnras,
283, 1361

\bibitem[ ] { } Blumenthal G.R., Faber S.M., Primack J.R., Rees M.J.,
1984, Nature, 311, 517

\bibitem[ ] { } Blanchard A., Valls--Gabaud D. \& Mamon G.A., 1992,
\aa, 264, 365

\bibitem[ ] { } Bond J.R. \& Efstathiou, 1984, \apj, 285, L45


\bibitem[ ] { } Bower R.G., Benson A.J., Baugh C.M., Cole S., Frenk
C.S., Lacey C.G., 2000, \mnras, submitted, astro-ph/0006109

Castander F.J. \& Couch W.J., 1994, \mnras, 268, 345

\bibitem[ ] { } Bower R.G., Lucey, J.R., \& Ellis R.S., 1991,
\mnras, 254, 601

\bibitem[ ] { } Bryan, G.L., 2000, \apj, 544, 1

\bibitem[ ] { } Bryan, G.L. \& Norman, M.L. 1998, \apj, 495, 80.


\bibitem[ ] { } Cavaliere A., Menci N. \& Tozzi P., 1999, \mnras,
308, 599

\bibitem[ ] { } Cayatte V., Kotanyi C., Balkowski C. \& van
Gorkom J.H., 1994, \apj, 107, 1003 

\bibitem[ ] { } Chamaraux P., Balkowski C. \& G\'{e}rard E.,
1980, \aa, 83, 38

\bibitem[ ] { } Cole S., 1991, \apj, 367, 45


\bibitem[ ] { } David L.P., Forman W. \& Jones C., 1991, \apj,
369, 121

\bibitem[ ] { } David L.P., Nulsen P.E.J., McNamara B.R., Forman
W., Jones C., Ponman T., Robertson B. \& Wise M., 2000, in prep, astro-ph/0010224

\bibitem[ ] { } David L.P., Slyz A., Jones C., Forman W., Vrtilek
S.D. \& Arnaud K.A., 1993, \apj, 412, 479

\bibitem[ ] { } Edge A.C. \& Stewert G.C., 1991, \mnras, 252, 414

\bibitem[ ] { } Edge A.C., Stewert G.C., Fabian A.C. \& Arnaud K.A.,
1990, \mnras, 245, 559

\bibitem[ ] { } Efstathiou G., Davis M., Frenk C.S. \& White S.D.M.,
1985, \apjs, 57, 241.

\bibitem[ ] { } Eke V.R., Navarro J.F. \& Frenk C.S., 1998,
\apj, 503, 569

\bibitem[ ] { } Evrard A.E., 1988, \mnras, 235, 911

\bibitem[ ] { } Evrard A.E. \& Henry J.P., 1991, \apj, 383, 95


\bibitem[ ] { } Fabian A.C., Crawford C.S., Edge A.C. \& Mushotzky
R.F., 1994, \mnras, 267, 779

\bibitem[ ] { } Fairley B.W., Jones L.R., Scharf C., Ebeling H.,
Perlman E., Horner D., Wegner G. \& Malkan M., 2000, \mnras, 315, 669

\bibitem[ ] { } Finoguenov A., Reiprich T.H. \& B\"{o}hringer H.,
2000, A\&A, submitted, astro-ph/0010190

\bibitem[ ] { } Frenk C.S., White S.D.M., Bode P., Bond J.R., Bryan
G.L., Cen R., Couchman H.M.P., Evrard A.E., Gnedin N., Jenkins A.,
Khokhlov A.M., Klypin A., Navarro J.F., Norman M.L., Ostriker J.P.,
Owen J.M., Pearce F.R., Pen U.--L., Steinmetz M., Thomas P.A.,
Villumsen J.V., Wadsley J.W., Warren M.S., Xu G., Yepes G., 1999,
\apj, 525, 554


\bibitem[ ] { } Girardi M., Giuricin G., Mardirossian F.,
Mezzetti M. \& Boschin W., 1998, \apj, 505, 74

\bibitem[ ] { } Gunn K.F. \& Thomas P.A., 1996, \mnras, 281, 1133

\bibitem[ ] { } Haiman Z., Mohr J.J. \& Holden G.P., 2000, \apj,
544, in press, astro-ph/0002336

\bibitem[ ] { } Hjorth J., Oukbir J. \& van Kampen E., 1998,
\mnras, 298, L1

\bibitem[ ] { } Holder G.P., Mohr J.J., Carlstrom J.E., Evrard
A.E. \& Leitch E.M., 2000, \apj, 544, 629

\bibitem[ ] { } Horner D.J., Mushotzky R.F. \& Scharf C.A., 1999,
\apj, 520, 78

\bibitem[ ] { } Jing Y.P. \& Suto Y., 2000, \apj, 529, L69

\bibitem[ ] { } Jones, C \& Forman W., 1984, \apj, 276, 38

\bibitem[ ] { } Kaiser N., 1986, \mnras, 222, 323

\bibitem[ ] { } Kaiser N., 1991, \apj, 383, 104

\bibitem[ ] { } Kauffmann G., White S.D.M. \& Guiderdoni B.,
1993, \mnras, 264, 201

\bibitem[ ] { } Kuntschner H., 2000, \mnras, 315, 184

\bibitem[ ] { } Lloyd-Davies E.J., Ponman T.J., Cannon D.B., 2000,
\mnras, 315, 689

\bibitem[ ] { } Loewenstein M. \& Mushotzky R.F., 1996, \apj,
471, L83

\bibitem[ ] { } Markevitch M., 1998, \apj, 504, 27

\bibitem[ ] { } Mathiesen B. \& Evrard A.E., 2001, \apj, 546, 100

\bibitem[ ] { } Mathiesen B., Evrard A.E., Mohr J.J., 1999,
\apj, 520, L21

\bibitem[ ] { } Matsumoto H., Tsuru, T.G., Fukazawa Y., Hattori
M. \& Davis D.S., 2000, \pasj, 52, 153

\bibitem[ ] { } McNamara B.R., Wise M., Nulsen P.E.J., David
L.P., Sarazin C.L., Bautz M., Markevitch M., Vikhlinin A., Forman
W.R., Jones C. \& Harris D.E., 2000, \apj, 534, L135

\bibitem[ ] { } Metzler C.A., 1995, Ph.D. Thesis


\bibitem[ ] { } Metzler C.A. \& Evrard A.E., 1997, astro-ph/9710324

\bibitem[ ] { } Metzler C.A., White M., Norman M. \& Loken C.,
1999, \apj, 520, L9

\bibitem[ ] { } Mohr J.J. \& Evrard A.E.,  1997, \apj, 491, 38

\bibitem[ ] { } Mohr J.J., Mathiesen B. \& Evrard A.E., 1999,
\apj, 517, 627

\bibitem[ ] { } Mohr J.J., Reese E.D., Ellington E., Lewis A.D.,
Evrard A.E., 2000, \apj, 544, 109

\bibitem[ ] { } Mushotzky R.F. \& Loewenstein M., 1997, \apj,
481, L63

\bibitem[ ] { } Mushotzky R.F. \& Scharf C.A., 1997, \apj, 482, L13

\bibitem[ ] { } Nagai D., Sulkanen M.E. \& Evrard A.E., 2000,
\mnras, 316, 120

\bibitem[ ] { } Navarro J.F., Frenk C.S., White S.D.M., 1995,
\mnras, 275, 720

\bibitem[ ] { } Neumann D.M. and Arnaud M., 1999, \aa, 348, 711

\bibitem[ ] { } Nevalainen J., Markevitch M., Forman W., 1999, \apj,
528, 1

\bibitem[ ] { } Nevalainen J., Markevitch M., Forman W., 2000,
\apj, 532, 694

\bibitem[ ] { } Owen F., 2000, IAUJD, 10, 7

\bibitem[ ] { } Peebles P.J.E., 1982, \apj, 263, L1

\bibitem[ ] { } Pen, U.--L., 1999, \apj, 510, L1

\bibitem[ ] { } Pierre M., Bryan G., Gastaud R., 2000, \aa, 356, 403

\bibitem[ ] { } Rees M.J. \& Ostriker J.P., 1977, \mnras, 179, 541

\bibitem[ ] { } Roettiger K., Burns J.O. \& Loken C., 1996, \apj,
473, 651

\bibitem[ ] { } Somerville R.S., Primack J.R., 1999, \mnras,
310, 1087

\bibitem[ ] { } Stanford S.A., Holden B., Rosati P., Tozzi P.,
Borgani S., Eisenhardt P.R. \& Spinrad H., \apj, accepted, astro-ph/0012250

\bibitem[ ] { } Steidel C.C., Adelberger K.L., Giavalisco M.,
Dickinson M. \& Pettini M., 1999, \apj, 519, 1

\bibitem[ ] { } Tamara T., Kaastra J.S., Peterson J.R., Paerels
F., Mittaz J.P.D., Trudolyubov S.P., Stewart G., Fabian A.C.,
Mushotzky R.F., Lumb D.H. \& Ikebe Y., 2000, submitted, astro-ph/0010362

\bibitem[ ] { } Theuns, T., Mo, H.J. \& Schaye, J. 2000, MNRAS,
submitted, astro-ph/0006065

\bibitem[ ] { } Thomas P.A., Muanwong O., Pearce F.R., Couchman
H.M.P., Edge A.C., Jenkins A. \& Onuora L., 2000, \mnras,
submitted, astro-ph/0007348


\bibitem[ ] { } Tozzi P. and Norman C., 2000, \apj, submitted, astro-ph/0003289

\bibitem[ ] { } Trager S.C., Faber S.M., Worthey G. \&
Gonz\'{a}lez J.J., 2000, \apj, 120, 165

\bibitem[ ] { } Van Dokkum P.G., Franx M., Fabricant D., Kelson
D.D. \& Illingworth G.D., 1999, \apj, 520, L95

\bibitem[ ] { } White R.E., 1991, \apj, 367, 69

\bibitem[ ] { } White S.D.M. \& Frenk C.S., 1991, \apj, 379, 52

\bibitem[ ] { } White S.D.M. \& Rees M.J., 1978, \mnras, 183, 341

\bibitem[ ] { } Wu K.K.S., Fabian A.C., Nulsen P.E.J., 1998,
\mnras, 301, L20

\bibitem[ ] { } Wu K.K.S., Fabian A.C., Nulsen P.E.J., 1999a,
\mnras, submitted, astro-ph/9907112

\bibitem[ ] { } Wu K.K.S., Fabian A.C., Nulsen P.E.J., 1999b,
\mnras, submitted, astro-ph/9910122

\bibitem[ ] { } XSPEC User's Manual:\newline http://legacy.gsfc.nasa.gov/\-docs/\-xanadu/\-xspec/\-index.html

\end{thebibliography}
\end{document}